\documentclass[12pt]{article}

\usepackage{amssymb}
\usepackage{amsmath}
\usepackage{amsbsy}

\newcommand{\no}{\noindent}
\newcommand{\be}{\begin{equation}}
\newcommand{\ee}{\end{equation}}

\def\bs{\bigskip}

\def\A{\mathcal A}

\def\c{\mathcal C}

\def\E{\mathcal E}
\def\F{\mathcal F}

\def\H{\mathcal H}

\def\O{\mathcal O}

\def\r{\mathcal R}

\def\U{\mathcal U}
\def\V{\mathcal V}

\def\C{\mathbb C}

\def\m{\mathbb M}
\def\P{\mathbb P}
\def\R{\mathbb R}
\def\T{\mathbb T}
\def\u{\mathbb U}
\def\Z{\mathbb Z}

\def\i{\rm i}

\title{Wick rotation and the positivity of energy in quantum field theory}
\author{Maxim Kontsevich, I.H.E.S., Bures-sur-Yvette\\and \\Graeme Segal, All Souls College, Oxford}
\date{}

\begin{document}

\maketitle

\section {Introduction}

In conventional quantum theory the states of a system are represented by the rays in a complex Hilbert space $\H$, and the time-evolution is given by a one-parameter group of unitary operators
$$U_t \ = \ e^{{\rm i} Ht} : \H \to \H $$
(for $t \in \R$), generated by an unbounded self-adjoint operator $H$ called the {\it Hamiltonian}. Positivity of the energy corresponds to the fact that $H$ is {\it positive-semidefinite}, i.e. that the spectrum of $H$ is contained in $\R_+$. This is clearly equivalent to saying that the operator-valued function $t \mapsto U_t$ is the boundary-value of a  {\it holomorphic} function $t \mapsto U_t$ which is defined in the upper half-plane
$$\{t \in \C : {\rm Im}(t) > 0 \}$$
and is bounded in the operator norm.\footnote{The physically relevant condition is actually that the energy is bounded below: replacing the Hamiltonian $H$ by $H-c$ makes no observable difference. Rather than asking for $U_t$ to be bounded for Im$(t) > 0$ we could require  $||U_t|| \leq {\rm e}^{c{\rm Im}(t)}$ for some $c$.}

The holomorphic formulation helps us  see what a strong constraint the positivity of energy is. The boundary value of a bounded holomorphic function in the upper half-plane  must vanish identically if it vanishes on an open interval of the real axis, and so, if taken literally, positive energy implies that a state $\xi \in \H$ for which $U_t(\xi)$ belongs to a closed subspace $\H_0$ of $\H$ for all $t< 0$ must remain in $\H_0$ for all $t \geq 0$, i.e. ``nothing can happen for the first time"  --- a paradox pointed out by Fermi  as early as 1932 [F].

How can this notion be adapted to the context of quantum field theory? The essential feature of quantum field theory is that the observables of the theory are organized by their positions in a given space-time $M$, which we shall take to be a smooth $d$-dimensional manifold with a Lorentzian metric $g = (g_{ij}).$ We expect that energy, and its positivity, should also have a local aspect, which is encoded, for any quantum field theory, in the \emph{energy-momentum tensor}, its most basic local observable.

In the usual formulations of quantum field theory such as [SW], for each space-time point $x \in M$ there is a topological vector space $\O_x$ of observables at $x$, and the $\O_x$ fit together to form a vector bundle on $M$. The content of the theory is completely encoded\footnote{This is an oversimplification just for this introduction. In a gauge theory, for example, an observable such as a ``Wilson loop" --- the holonomy of the gauge field around a closed loop in space-time --- is localized not at a point but at a loop, and we shall not  exclude such features.} in multilinear `maps'
\begin{eqnarray*}
\O_{x_1} \times \ldots \times \O_{x_k} \ \ & \longrightarrow & \ \ \ \C \ \ \ \ \ \ \ \ \ \ \ \ \ \ \ \ \ \ \ \ \ \ \ \ \ \ \ \ \ \ \ \ \ \  \ \ \ \ \ (1) \\
(f_1, \ldots , f_k) \ \ \ \ & \mapsto & \ \ \langle f_1, \ldots , f_k \rangle_{(M,g)}
\end{eqnarray*}
 for all sequences $\{x_1, \ldots , x_k\}$ of points in $M$, defining generalized functions\footnote{The Wightman axioms ask for the vacuum expectation values to be distributions on $M^k$ (which morally means that the theory has a logarithmic-conformal limit at short distances), but our formulation will not exclude such examples such as the sigma-model with a circle as target, for which, when $d \geq 3$, the vacuum expectation values are hyperfunctions but not distributions.}  on the products $M^k$. The functions (1) are called {\it vacuum expectation values}. To come from a field theory they must satisfy a long list of conditions such as the Wightman axioms given in [SW]. These include a \emph{causality} axiom which asserts that if the points $x_1,\ldots , x_k$ are spatially separated (i.e. no two can be joined by a path whose speed is never faster than light) then the expectation value is independent of the ordering of the points.

One motivation for this formulation is the ``path-integral" picture (cf. [FH]), according to which the theory arises from a mythological superstructure consisting of a space $\Phi_M$ of ``fields" of some kind which are {\it locally defined} on the Lorentzian manifold $(M,g)$. In this picture the vector space $\O_x$ of observables at $x$ is the space of smooth functions $f:\Phi_M \to \C$ such that $f(\phi)$ depends only on the restriction  of $\phi$ to an arbitrarily small neighbourhood of $x$.  All of the physics of the theory is determined by an {\it action functional} $S_g: \Phi_M \to \R$ which notionally defines a complex-valued measure on the space $\Phi_M$, symbolically denoted by ${\rm e}^{-{\rm i}S_g(\phi)/\hbar}\mathcal D \phi$ 
. The parameter $\hbar$ here --- the unit of action --- is Planck's constant.  The vacuum expectation values are given in terms of the measure by
$$\langle f_1, \ldots ,f_k \rangle_{(M,g)} \ = \ \int_{\Phi_M}f_1(\phi) \ldots f_k(\phi) {\rm e}^{-{\rm i}S_g(\phi)/\hbar}\mathcal D\phi.$$

The smallness of the unit $\hbar$ of action means that the notional integral is very highly oscillatory, and so the measure on $\Phi_M$ is effectively concentrated near the critical points of the action. These points are the solutions of the classical equations of motion, and they form the classical state space of the system. 

\bs

There are two ways to introduce the idea of positive energy into this picture. Both involve holomorphicity, and we shall refer to both --- rather vaguely --- as `Wick rotation'. They derive from two different ways of viewing the time $t$ in the evolution-operator $U_t$ of quantum mechanics. The traditional way is to regard the possibility of extending the map $t \mapsto U_t$ to the upper half-plane as ``creating" a complex time-manifold with the physical time-axis at its boundary. In field theory this leads to viewing space-time $M$ as part of the boundary of a complex manifold $M_{\C}$, and then the positivity of energy is expressed by the property that the the vacuum expectation values (1) are the boundary-values of holomorphic functions on a domain in $(M_{\C})^k$. This makes good sense when $M$ is the standard Minkowski space $\m \cong \R^{3,1}$. It is less natural in the case of a curved space-time, if only because a smooth manifold does not have a complexification (or even a way of putting it on the boundary of a complex manifold) until one chooses --- non-canonically --- a real-analytic structure on it.  Even then, $M$ may have only a small thickening as a complex manifold, while the holomorphic characterization of positive energy makes use of the {\it whole} upper half of the $t$-plane.

An alternative approach --- the one we present in this paper --- is to treat the time-parameter $t$  as the {\it length} of an oriented time-interval, thinking of it as a 1-dimensional manifold  equipped with a Riemannian (or pseudo-Riemannian) metric. Then we do not need to complexify the time-manifold: we simply allow the {\it metric} on it to be complex-valued. (The authors have independently spoken about this idea from time to time since the late 1980s --- see Section 4 --- but as far as we know no-one else has pursued it systematically.) There are two reasons why the approach fits well with the path-integral picture when the time-interval of quantum mechanics is replaced by the space-time $M$ of quantum field theory. First, the usual action-functionals $S_g$ depend explicitly on the Lorentzian metric $g$ of $M$ in a way that makes sense when $g$ is complex. Secondly and more importantly, the path-integral is an \emph{oscillatory} integral which does not converge even schematically. Its archetype is an improper Gaussian integral of the form
$$F(A) \ = \ \int_{\R^n}\exp \left(\frac{{\rm i}}{2}x^T A x\right) \  dx_1 \ldots dx_n, \ \ \ \ \ \ \ \ \ \ \ \ \ \ \ \ \ \ \ (2)$$
where $A$ is a real symmetric $n \times n$ matrix. The standard way to treat such an integral is to begin with a complex symmetric matrix $A$ whose imaginary part is positive definite --- i.e.  a point $A$ of the Siegel `generalized upper half-plane'. For such  matrices the integral converges and defines a holomorphic function of $A$ in the Siegel domain. The value of the original improper integral is defined as the limit  as $A$ moves to the boundary of the domain.

The main point of the present paper is to introduce an interesting domain Met$_{\C}(M)$ of complex-valued metrics on a smooth manifold $M$. The positivity of energy of a quantum field theory is expressed by the property that it is defined for space-time manifolds with metrics belonging to this domain. The domain is a complexification of the manifold Met$(M)$ of ordinary Riemannian metrics on $M$, and the real Lorentzian metrics (but \emph{not} real metrics of other signatures) are a  subset of its boundary. The special role of Lorentzian signature is perhaps the most notable feature of our work. In Section 5 we shall explain how a theory defined on space-times with complex metrics gives rise, under appropriate conditions, to a theory defined for Lorentzian space-times which automatically satisfies the expected causality axiom when the Lorentzian metric is globally hyperbolic.   Finally, although we avoid complexifying space-time, our approach  leads us  to a conjecture about a question arising in the traditional treatment of quantum field theories defined in Minkowski space $\m$:  how to characterize the largest domain in $(\m_{\C})^k$ in  which the vacuum expectation values are holomorphic. 

\bs

\newpage

\no{\bf The Shilov boundary}

\bs

The relevant meaning of `boundary' for the complex domains we are interested in is the  {\it Shilov boundary}. An account of this concept can be found in [H\"or], but our use of it will be little more than heuristic. It is the analogue for a complex domain of the `extremal points' of a bounded open subset $U$ of $\R^n$. In the Euclidean situation the extremal points are the smallest subset $K$ of the closure of $U$ such that every affine-linear  function on $U$ attains its maximum on $K$, and the \emph{convex hull} of $U$ is the set of points $x \in \R^n$ where 
$$ \inf_K f \ < \ f(x) \ < \ \sup_K f$$  for  every affine function $f$. For a complex domain, we replace affine-linear functions by holomorphic functions.   Thus if $U$ is an open subset of a finite-dimensional complex Stein manifold $U^+$ (e.g. an affine space or an affine algebraic variety), and the closure of $U$ in $U^+$ is a compact manifold $X$ with a piecewise-smooth boundary, then the Shilov boundary of $U$ is the smallest\footnote{A short proof that there is a unique such smallest subset can be found in [H\"or] p.67.} compact subset $K$ of $X$ with the property that for every holomorphic function $f$ defined in a neighbourhood of  $X$ we have
$$ \sup_U |f| \ = \ \sup_K |f|.$$
 The Shilov boundary of a manifold is part of its topological boundary, but can be much smaller, just as, for example, the only extremal points of a Euclidean simplex are its vertices. In our examples its real dimension will always be equal to the complex dimension of the domain. Thus, for the polydisc 
$$U \ = \ \{(z_1, \ \ldots \ ,z_n) \in \C^n : |z_i| < 1\},$$
the Shilov boundary is the torus $|z_1| = \ldots = |z_n| = 1$

 The most relevant example for us is the Siegel `generalized half-plane' $\u_n$ of complex-valued quadratic forms on $\R^n$ with positive-definite real part, i.e. the $n \times n$ complex symmetric matrices $A$ with Re$(A)$ positive-definite.  As so presented, $\u_n$ is not bounded in the vector space of matrices, but it has an alternative ``unit disc" description as the complex symmetric matrices $A$ such that $||A|| < 1$, or, equivalently, such that $1 - \bar A A$ is positive definite. (The second description is obtained from the first by the \emph{Cayley transform} $A \mapsto (A-1)(A + 1)^{-1}$.) In the first description, the purely imaginary symmetric matrices lie on the boundary of the domain as the ``generalized imaginary axis". They form a dense open subset of the Shilov boundary, just as the imaginary axis is a dense open subset of the boundary of the usual right half-plane.  But to understand the Shilov boundary in this case it is better to pass to yet another description of $\u_n$, as  an open subset of the compact complex manifold Lag$(\C^{2n})$ of complex Lagrangian subspaces of a symplectic vector space $\C^{2n}$.  

To obtain this description, let us start from a real vector space $V$ with complexification $V_{\C}$. Complex-valued quadratic forms on $V$ are the same as symmetric maps $A: V_{\C} \to V_{\C}^*$, and the graph of such a map is a Lagrangian subspace of the complex symplectic vector space $V_{\C} \oplus V_{\C}^*$. Now any Lagrangian subspace $W$ of $V_{\C} \oplus V_{\C}^*$ acquires a Hermitian inner product by the formula $\langle w_1,w_2\rangle = {\rm i}S(\bar w_1, w_2)$, where $S$ is the $\C$-bilinear symplectic form  of $V_{\C} \oplus V_{\C}^*$.  The Siegel domain $\u(V)$ consists precisely of those $W$ for which the Hermitian form of $W$ is positive-definite. The topological boundary of the domain consists of all $W$ whose Hermitian form is positive-\emph{semidefinite} but not positive-definite.  It is a piecewise-smooth manifold stratified by the rank of the Hermitian form. The lowest-dimensional stratum, where the Hermitian form vanishes, is the smooth compact manifold Lag$(V \oplus V^*)$ of real Lagrangian subspaces of $V \oplus V^*$. It has a dense open subset consisting of subspaces which do not intersect $V^*$: these are the graphs of the real symmetric maps $A:V \to V^*$. 

In this example, and all the others we shall encounter, there is no difficulty in \emph{identifying} the Shilov boundary inside the topological boundary, for by the maximum-modulus principle a point does \emph{not} belong to it if it can be written as $f(0)$ for some non-constant holomorphic map $f$ from a neighbourhood of $0$ in $\C$ to the closure of the domain. In particular, we shall meet \emph{tube domains} of the form $\R^N \times {\rm i}C \subset \C^N$, where $C$ is a convex open subset of $\R^N$: for them the Shilov boundary is $\R^N \times \i K$, where $K$ is the set of extremal points of $C$.

\section{The domain of complex metrics}

A Riemannian metric on a manifold $M$ is a positive-definite symmetric bilinear form $g: T_x \times T_x \to \R$ on the tangent space $T_x$ at each point $x \in M$. The metrics we shall consider will be defined by symmetric $\R$-bilinear maps $g: T_x \times T_x \to \C$ at each point, with an appropriate generalization of the positivity condition.

To see what condition we should require, let us consider the simplest example of a field theory: a free real scalar field of mass $m$. Then the space of `fields' $\Phi_M$ is the vector space C$^{\infty}(M;\R)$ of smooth functions, and in the exponent of the path-integral we have the quadratic form
\begin{eqnarray*}
{\rm i} S_g(\phi)  \ \  &=& \  \ \frac{1}{2}\int_M ( d\phi\wedge*d\phi + m^2 \phi\wedge *\phi) \\ &=& \  \ \frac{1}{2}\int_M \left\{\sum g^{ij}\frac{\partial \phi}{\partial x^i}\frac{\partial \phi}{\partial x^j} + m^2\phi^2 \right\} ( \det g)^{1/2} |dx^1 \ldots dx^d|.
\end{eqnarray*}
Here $(g^{ij})$ denotes the inverse of the matrix $g = (g_{ij})$, and $*$ is the Hodge star-operator  defined by the metric, which takes differential forms of degree $p$ to  forms of degree $d-p$ twisted by the orientation bundle.  (We shall not assume the space-time $M$ is orientable.) In particular the star-operator takes the constant function 1 to the volume element 
$$*1 \ = \ {\rm vol}_g \ = \ ({\rm det} g)^{1/2}|dx^1 \ldots dx^d| \ \ \ \ \ \ \ \ \ \ \ \ \ \ \ \ \ \ \ \ \ \ \ \ \ (3)$$
Notice that for a Lorentzian metric $g$ the volume element $*1$ is pure imaginary. This agrees with the fact that the `action' $S_g$ should be real for a Lorentzian manifold. We want the real part of the quadratic form i$S_g$ to be positive-definite for all the complex metrics we allow. This imposes two conditions. First,  we need the real part of the twisted $d$-form vol$_g$ defined by the formula (3) to be a positive volume-form on $M$. We therefore require that $\det g$, which is invariantly defined up to multiplication by a positive real number, is \emph{not} real and negative, and  we choose $(\det g)^{1/2}$ to have positive real part. 

The second condition we need is that the real part of the matrix (det$g)^{1/2}g^{-1}$ --- or equivalently of the inverse matrix $({\rm det}g)^{-1/2}g$ --- is positive-definite.  The two conditions together would give us a domain whose  Shilov boundary (like that of the Siegel generalized half-plane) contains indefinite real quadratic forms of all signatures, and not only the Lorentzian ones. But we shall impose further conditions. A clue to what more is needed comes from the theory of the electromagnetic field on $M$, with its field-strength given by a real 2-form $F$ on $M$, and with the action-functional
$${\rm i} S_g(F) \ = \ \frac{1}{2}\int_M F \wedge * F.$$
The Hodge $*$-operator makes sense for a complex metric: for a $p$-form $\alpha$ we define a twisted $(d-p)$-form $*\alpha$  by taking the inner-product of $\alpha$ with vol$_g \ = \ *1$, using the complex inner-product $g$.  We regard vol$_g$ as an element of the complex line $|\wedge^d(T^*_x)|_{\C}$, where $|\wedge^d(T^*_x)|$ is the tensor product of $\wedge^d(T^*_x)$ with the real orientation line of $T_x$, and $|\wedge^d(T^*_x)|_{\C}$ is its complexification, \emph{but} with the convention that the orientation-reversing automorphisms of $T_x$ act \emph{antilinearly}. We say that an element of the real part of the line is positive if it is a positive volume-element.

For the electromagnetic field we need the real part of the quadratic form
$$\wedge^2(T_x^*)  \ \longrightarrow \ |\wedge^d(T_x^*)|_{\C}$$ 
given by $F \mapsto F\wedge *F$ to be positive-definite. 

This makes it natural, if we are going to consider space-time manifolds $M$ of all dimensions, to propose 

\bs

\no{\bf Definition 2.1} \ \  {\it On a $d$-dimensional real vector space $V$ a quadratic form $g:V \to \C$ is called an {\rm allowable} complex metric if, for all degrees $p \geq 0$, the real part of the quadratic form 
$$\wedge^p(V^*) \ \longrightarrow \ |\wedge^d(V^*)|_{\C}$$
given by  $\alpha \mapsto \alpha \wedge * \alpha$ is positive-definite.} 

\bs

Fortunately, this definition has an equivalent formulation which is much more explicit and illuminating.

\bs

\no{\bf Theorem 2.2} \ \ {\it Definition 2.1 is equivalent to: there is a basis of the real vector space $V$ in which the quadratic form $g$ can be written
$$\lambda_1 y_1^2 + \lambda_2 y_2^2 + \ldots + \lambda_d y_d^2,$$
where the $y_i$ are coordinates with respect to the basis, and the $\lambda_i$ are non-zero complex numbers, not on the negative real axis, such that}
$$|\arg (\lambda_1) | + |\arg (\lambda_2) | + \ldots + |\arg (\lambda_d)| < \pi. \ \ \ \ \ \ \ \ \ \ \ \ \ \ \ (4)$$ 

\bs

The complex-valued quadratic forms  $g:V \to \C$ on a real vector space $V$ which satisfy the conditions of (2.1) or (2.2) form an open subset  $Q_{\C}(V)$ of the complex vector space $S^2(V^*_{\C})$.   It  follows from Theorem 2.2 that the real inner products with signature $(d-1,1)$ --- but not those with other signatures --- lie on the boundary of the domain $Q_{\C}(V)$. For if the metric is real then each $|\arg(\lambda_i)|$ is either 0 or $\pi$, and the inequality (4)  shows that at most \emph{one} of the $|\arg(\lambda_i)|$ can become $\pi$ on the boundary. 

Another consequence of (4) is that 
$$\max \arg \lambda_i \ - \ \min \arg \lambda_i \ < \ \pi,$$
which shows  that when $v$ runs through $V$ the complex numbers $g(v)$ form a closed convex cone in $\C$ disjoint from the open negative real axis. In particular, $g(v)$ can never be real and negative.
  
  Using the criterion mentioned at the end of Section 1 we see that the real Lorentzian metrics --- and no other nondegenerate metrics ---  belong to the \emph{Shilov} boundary of $Q_{\C}(V)$, when it is regarded as a bounded domain in an affine variety (cf. the proof of 2.7 below).  Indeed if $g = \sum \lambda_j y_j^2$ is a complex metric for which the inequality (4) becomes an equality, and at least two of the eigenvalues $\lambda_j$ and $\lambda_k$ are not on the negative real axis, then (after rescaling the basis vectors $e_j$ and $e_k$ so that $|\lambda_j | = |\lambda_k| = 1$) we get a holomorphic curve through $g$, in the closure of $Q_{\C}(V)$, by changing $\lambda_j$ to $(\lambda_j)^{1+z}$ and $\lambda_k$ to $(\lambda_k)^{1-\varepsilon z}$, where $\varepsilon$ is $+1$ or $-1$ according as the arguments of $\lambda_j$ and $\lambda_k$ have the same or opposite signs. 
  
  In fact the Shilov boundary of $Q_{\C}(V)$ contains \emph{two} disjoint copies of the space of Lorentzian metrics on $V$, for an eigenvalue $\lambda$ can approach the negative real axis either from above or from below. The two copies are interchanged by the complex-conjugation map on $Q_{\C}(V)$. Because of our choice to make the orientation-reversing elements of GL$(V)$  act antilinearly on the orientation-line of $V$, we can say that the nondegenerate points of the Shilov boundary of $Q_{\C}(V)$ are the \emph{time-oriented} Lorentzian metrics.

\bs

We define the space Met$_{\C}(M)$ of allowable complex metrics on a smooth manifold $M$ as the space of smooth sections of the bundle on $M$ whose fibre at $x$ is $Q_{\C}(T_x)$. 

\bs

Before giving the surprisingly simple proof of Theorem 2.2
let us say what motivated  the two different-looking  conditions. The desire to make the real parts of natural quadratic action functionals positive-definite hardly needs further comment, but choosing to focus on the  `higher abelian gauge field' actions $\alpha \wedge  * \alpha$ --- the `Ramond-Ramond' fields of superstring theory ---  may well seem arbitrary. Why not allow other kinds of tensor fields? Our conditions do not imply that they will be positive-definite. Witten has kindly suggested to us a justification for our focus, based on properties of the classical energy-mometum tensor explained in [WW]. Including the higher gauge theories does, in any case, impose an \emph{upper} bound on the class of complex metrics we can allow, for the  partition functions of these theories on a  $d$-dimensional torus $M$ with a flat Riemannian metric $g$ are explicitly known (cf. [Ke], [Sz](4.4)), and we can see to which complex metrics they can be analytically continued. The gauge-equivalence classes of fields form an infinite-dimensional Lie group which is a product of a torus, a lattice, and an infinite-dimensional real vector space, and the partition function is the product of three corresponding factors.  More precisely, an abelian gauge $(p-1)$-field $A$ has a \emph{field-strength} $F_A$, a closed $p$-form on $M$ with integral periods, which determines $A$ up to the finite-dimensional torus $H^{p-1}(M;\T)$ of flat gauge fields with $F_A = 0$. The space of fields is therefore a product
$$H^{p-1}(M;\T) \ \times \ \Phi_p \ \times \ \Gamma_p,$$
where $\Phi_p$ is the vector space of exact $p$-forms on $M$, and $\Gamma_p \cong {\rm Harm}_{\Z}^p(M) \cong H^p(M;\Z)$ is the finite-dimensional lattice of harmonic (and hence constant) $p$-forms with integral periods. The partition function is a Gaussian integral on this product: the torus of flat fields contributes its volume (for an appropriate metric determined by the geometry of $M$), the lattice $\Gamma_p$  of harmonic $p$-forms contributes its theta-function $$\sum_{\alpha \in \Gamma_p}  \ \exp \left(-\frac{1}{2} \int_M\ \alpha \wedge *\alpha \right ) \ ,$$
while the vector space $\Phi_p$ contributes an `analytic torsion' which is a power of the determinant of the Laplace operator acting on smooth functions on $M$ (with the zero-eigenvalue omitted) --- an analogue of the Dedekind eta-function, but with the lattice of characters of the torus $M$ replacing the lattice $\Z + \tau \Z \subset \C$. Of these three  factors, the first clearly extends holomorphically to the space of all flat complex metrics on $M$, and the analytic torsion can be continued to a non-vanishing holomorphic function in the open set of complex metrics $g$ for which (det  $g)^{-1/2}g$ belongs to the Siegel domain $\u(V))$; but the theta-function cannot be continued beyond those metrics for which the real part of the form $\int \alpha \wedge *\alpha$ is positive.

\bs

Approaching from the opposite direction, the inequality (4) is motivated by 
% \emph{lower} bound on the class of complex metrics we would like to include, coming from
the traditional analytical continuation of vacuum expection values to an open subset of the $k$-fold product of complexified Minkowski space $\m_{\C}$. The Wightman axioms imply that the expectation values extend holomorphically to a domain $\U_k$ called the `permuted extended tube'\footnote{A set of points $x_1, \ldots , x_k$ belongs to $\mathcal U_k$ if, after ordering them suitably, there is an element $\gamma$ of the complexified Lorentz group such that the imaginary part of $\gamma(x_i - x_{i+1})$ belongs to the forward light-cone for each $i$. }, which is functorially associated to $\m_{\C}$ with its $\C$-bilinear metric.  It is a basic result in the Wightman theory (cf. [SW], or [Ka](2.1)) that $\U_k$ contains the configuration space Conf$_k(\mathbb E)$ of all $k$-tuples of \emph{distinct} points of the standard Euclidean subspace $\mathbb E \subset \m_{\C}$. For a $d$-dimensional real vector space $V$ with a complex metric the complexification $V_{\C}$ is isomorphic to $\m_{\C}$, uniquely up to a complex Lorentz transformation, and so the domain $\U_k(V)$ is well-defined in $(V_{\C})^k$. In  the next section we shall give a definition of a quantum field theory on space-times $M$ with complex metrics: it implies that the expectation values are smooth functions on the configuration spaces Conf$_k(M)$ of distinct $k$-tuples in $M$.  That makes it natural to %include among the allowable metrics those of  all the $d$-dimensional real linear subspaces $V$ of $\m_{\C}$  such that
ask which (constant) complex metrics on $V$  have the property that the configuration space Conf$_k(V)$  is contained in the holomorphic envelope of $\U_k(V)$, i.e.  the largest Stein manifold to which all holomorphic functions on $\U_k(V)$ automatically extend. The original motivation of condition (4) was

\bs

\no{\bf Proposition 2.3} \ {\it If a complex metric on a $d$-dimensional real vector space $V$ satisfies condition (4)  then ${\rm Conf}_k(V)$ is contained in the holomorphic envelope of $\U_k(V)$.}

\bs

 We shall postpone the  proof of this result to an appendix at the end of this section. 

\bs
     
\no{\it Proof of Theorem 2.2} \ \  The first point is to show that a quadratic form which satisfies the conditions of Definition 2.1 can be written in the diagonal form $\sum \lambda_j y_j^2$ with respect to  real coordinates $y_j$ on $V$. To diagonalize a complex form $g = A+ {\rm i} B$ with respect to a real basis is to diagonalize its real and imaginary parts simultaneously, which is possible if either $A$ or $B$ --- or, more generally, a real linear combination of them such as the real part of $({\rm det}g)^{-1/2}g$ --- is positive-definite. But 2.1, applied when  $p=1$, implies that the real part of $({\rm det}g)^{-1/2}g$ is positive.

Suppose now that $g$ is diagonalized  with respect to a basis $\{e_i\}$ of $V$. Then the form $\alpha \mapsto \alpha \wedge * \alpha$ on $\wedge^p(V^*)$ is diagonal with respect to the basis $\{e_S^* = e^*_{i_1}\wedge \ldots \wedge e^*_{i_p}\}$, where $\{e_i^*\}$ is the dual basis to $\{e_i\}$, and $S$ runs through $p$-tuples $S = (i_1, \ldots , i_p)$. The value of the form  $\alpha \wedge * \alpha$ on the basis element $e^*_S$ is
$$(\lambda_1 \ldots \lambda_d)^{1/2}\prod_{i \in S}\lambda_i^{-1},$$
which has positive real part if its argument
$$\frac{1}{2}\left\{\sum_{i \in S} {\rm arg}(\lambda_i) - \sum_{i \not \in S} {\rm arg}(\lambda_i)\right\}$$
lies in the open interval $(-\pi/2,\pi/2)$. But to say that this is true for every subset $S$ of $\{1, \ldots , d\}$ is precisely condition $(4)$.  $\spadesuit$

\bs

The proof of Theorem 2.2 shows that to give an element $g$ of $Q_{\C}(V)$ is the same as to give a finite sequence $\theta_1 \geq \theta_2 \geq \ldots \geq \theta_m$ in the interval $(-\pi, \pi)$ together with a  decomposition
 $$V = V_1 \oplus \ldots \oplus V_m$$
 such that
 $$\sum_k \dim V_k\cdot |\theta_k|<\pi. $$ Thus on $V_k$ the bilinear form $g$ is ${\rm e}^{{\rm i} \theta_k}$ times a real positive-definite form.  The only ambiguity in this description is that if, say, $\theta_k = \theta_{k+1}$ we can  replace $V_k$ by $V_k \oplus V_{k+1}$ and omit $\theta_{k+1}$ and $V_{k+1}$. This means that the subspace $P = \bigoplus {\rm e}^{-{\rm i} \theta_k/2}V_k$ of the complexification $V_{\C}$ of $V$ is \emph{canonically} associated to the form $g$. On the real subspace $P$ the complex bilinear form $g$ is real and positive-definite. Our argument  gives us  canonical isomorphisms 
 $$V \ = \ \exp({\rm i}\pi\Theta/2)(P) \ \subset \ P_{\C} \ = \ V_{\C},$$
 where $\Theta:P \to P$ is the self-adjoint operator which is multiplication by $\theta_k$ on $P_k = {\rm e}^{-{\rm i}\theta_k/2} V_k$. Condition $(4)$ becomes the assertion that $\Theta$ has trace-norm\footnote{The trace-norm is the sum of the absolute values of the eigenvalues.} $||\Theta ||_1 < 1$. 
This shows that the space $Q_{\C}(V)$ is parametrized by the pairs $(g_0, \Theta)$, where $g_0$ is a positive-definite inner-product on $V$ and $\Theta$ belongs to the  convex open set $\Pi(V,g_0)$ of operators in $V$ which are self-adjoint with respect to $g_0$ and satisfy $||\Theta||_1 < 1$, i.e. the interior of the convex hull of the rank 1 orthogonal projections in $V$.  In fact we have proved
 
 \bs
 
 \no{\bf Proposition 2.4} \ \ $Q_{\C}(V)$ {\it is a fibre-bundle over the space of positive-definite inner products on $V$ whose fibre at a point $g_0$ is $\Pi(V,g_0)$. Equivalently, choosing a reference inner-product on $V$, we have
 $$Q_{\C}(V) \ \cong \ {\rm GL}(V) \times_{{\rm O}(V)} \Pi(V).$$
 In particular, $Q_{\C}(V)$ is contractible.}
 
 \bs

It is an important fact that an allowable complex metric on $V$ remains allowable when restricted to any subspace $W$ of $V$. This follows from an analogous property of the trace-norm, but we shall give a direct proof, as its point of view on the angles $\theta_i$ as critical values  helps give a feeling for allowable complex metrics.

\bs 

 \no{\bf Proposition 2.5} \  \ {\it If $g \in Q_{\C}(V)$ and $W$ is any vector subspace of $V$ then $g|W$ belongs to $Q_{\C}(W)$. }
 
 \bs

\no{\it Proof} \ \ \  For any $g \in Q_{\C}(V)$ the function $v \mapsto {\rm arg}(g(v))$ is a smooth map from the real projective space $\P(V)$ to the open interval $(-\pi, \pi) \subset \R$.  By rescaling the basis elements $\{e_k\}$  we can write $g$ as $\sum{\rm e}^{{\rm i}\theta_k}y_k^2$. The numbers $\theta_k$ are precisely the critical values of arg$(g)$.  We shall order the basis elements so that
$$\pi \ > \ \theta_1 \ \geq \ \theta_2 \ \geq \ \ldots \ \geq \ \theta_d \ > \ -\pi.$$

For each vector subspace $A$ of $V$  let us write $\theta^A$ and $\theta_A$ for the supremum and infimum of arg$(g)$ on $\P(A)$.  Then we have
$$\theta_k  \ \ = \ \ {\rm sup}\{\theta_A: {\rm dim}(A) = k\} \ \ = \ \ {\rm inf}\{\theta^A: {\rm dim}(A) = d-k+1\}.$$
It is enough to prove Proposition 2.5 when $W$ is a subspace of $V$ of codimension 1. In that case the preceding characterization of the critical values shows that if 
$\theta'_1 \geq \ldots \geq \theta'_{d-1}$
are the critical values of arg$(g|W)$  we have $\theta_k \geq \theta'_k \geq \theta_{k+1}$. The critical values for $g|W$ therefore \emph{interleave} those for $g$:
$$\theta_1 \geq \theta'_1 \geq \theta_2 \geq \theta'_2 \geq \ldots \geq \theta_{d-1} \geq \theta'_{d-1} \geq \theta_d.$$
This implies that $\sum |\theta'_k| \leq \sum |\theta_k| < \pi$, as we want. $ \ \spadesuit$

\bs

In Section 5 we shall need the following variant of the preceding formulation. 
Suppose that $Z$ is a $d$-dimensional complex vector space with a nondegenerate quadratic form $g$. (Any such pair $(Z,g)$ is isomorphic to $\C^d$ with the standard form $\sum z^2_k$.) Let $\r(Z)$ denote the space of all $d$-dimensional real subspaces $A$ of $Z$ such that $g|A$ belongs to $Q_{\C}(A)$. This is an open subset of the Grassmannian of all real subspaces of $Z$. If $Z_{\R}$ is any $d$-dimensional real vector subspace of $Z$ for which $g|z_{\R}$ is real and positive-definite then  the projection $A \subset Z \to Z_{\R}$ is an isomorphism, for any non-zero element of its kernel would have the form i$v$ with $v \in Z_{\R}$, and so $g({\rm i}v)$ would be real and negative, which cannot happen if $g|A$ is allowable.

\bs

\no{\bf Proposition 2.6}  \ \ {\it The space $\r(Z)$ is contractible, and is  isomorphic to}
$$ {\rm O_{\C}}(Z) \times_{{\rm O}(Z_{\R})} \Pi(Z_{\R}).$$

\bs

\no{\it Proof} \ \ This is essentially a reformulation of what has been said, but it may be helpful to relate the spaces $Q_{\C}(V)$ and $\r(Z)$ by considering, for a complex quadratic vector space  $(Z,g)$ as above, the intermediate space $\r(V;Z)$ of $\R$-linear embeddings $f:V \to Z$ of the real vector space $V$  such that $f^*(g)$ is allowable. This space has two connected components, corresponding to the orientation of the projection $V \to Z_{\R}$.

The groups GL$(V)$ and O$_{\C}(Z)$ act by right- and left-composition on $\r(V;Z)$, and each action is free. Thus $\r(V;Z)$ is at the same time a principal GL$(V)$-bundle with base $\r(Z)$ and a principal O$_{\C}(Z)$-bundle with base $Q_{\C}(V)$. But the Lie groups GL$(V)$ and O$_{\C}(Z)$ are homotopy equivalent to their maximal compact subgroups, i.e. in both cases to the compact orthogonal group O$_d$. More precisely, the contractibility of $Q_{\C}(V)$ implies that $\r(V;Z)$ is homotopy-equivalent to the fibre O$_{\C}(Z)f$ for any  $f\in \r(V;Z)$. If we choose $f$ so that $f^*(g)$ is a positive-definite real form on $V$ this gives us a homotopy-equivalence $ {\rm O}(V) \to {\rm O}_{\C}(Z)f \to \r(V;Z)$. But O$(V)$ is also contained in and equivalent to the fibre $f{\rm GL}(V)$ of the other fibration $\r(V;Z) \to \r(Z)$, which implies the contractibility of its base $\r(Z)$. \ $\spadesuit$

\bs

The last property of $Q_{\C}(V)$ which we shall record briefly, for the sake of experts, is

\bs

\no{\bf Proposition 2.7} \  \ {\it The domain $Q_{\C}(V)$ is holomorphically convex, i.e.\ a `domain of holomorphy'.}

\bs

\no{\it Proof} \ \ The Siegel domain $\u(V)$ of  complex-valued inner products with positive-definite real part on a real vector space $V$ is known to be a domain of holomorphy in $S^2(V^*_{\C})$. So therefore is the product
$$\prod_{0 \leq p \leq d/2} \u(\wedge^p(V))$$
inside its ambient complex vector space. The space $Q_{\C}(V)$ is the intersection of this product domain with the affine variety which is the natural embedding of $S^2(V^*_{\C})$ in this ambient vector space, and so it too is a domain of holomorphy. \ $\spadesuit$

\bs

\bs
 
\no {\bf The two-dimensional case}

\bs

The case $d = 2$ is especially simple because then the matrix $(\det g)^{- 1/2}g$ depends only on the conformal structure, and decouples from  the volume element.

A non-degenerate complex inner product $g$ on a 2-dimensional real vector space $V$ is determined up to a scalar multiple by its two distinct null-directions in the complexified space $V_{\C}$. We can think of these as two points of the Riemann sphere $\P(V_{\C})$. Then  $(\det g)^{-1/2}g$ has positive real part precisely when the two points lie one in each of the open hemispheres of the sphere $\P(V_{\C})$ separated by the real equatorial circle $\P(V)$. When the two points move to distinct points of the equator we get a Lorentzian inner product, with its two light-directions in $\P(V)$.

A point of the sphere $\P(V_{\C})$ not on the equator can be regarded as a complex structure on the real vector space $V$, and the two hemispheres correspond to the two possibilities for the orientation which a complex structure defines. On a smooth surface $\Sigma$ any almost-complex structure is integrable, so a point of Met$_{\C}(\Sigma)$ is a pair of complex structures on $\Sigma$ of opposite orientations, together with a complex volume element. The Riemannian metrics are those for which the two complex structures are complex-conjugate to each other, and the volume element is real.

When $d=2$ the domain $Q_{\C}(V)$ is thus a 3-dimensional polydisc, one disc for each of the complex structures, and the third for the volume-element.

\bs

\newpage

\no{\bf The one-dimensional case: electric circuits}

\bs

Our concept of an allowable complex metric does not at first look interesting in the one-dimensional case, but if we allow \emph{singular} 1-manifolds --- identified with finite graphs $M$ --- we find that complex metrics arise naturally in electrical circuit theory. A Riemannian metric on $M$ is determined (up to isometry) by the assignment of a positive real number to each edge of the graph, and can be interpreted as its \emph{resistance} when the edge is regarded as a wire in an electrical circuit. A state of the system (perhaps with current entering or leaving at each node) is determined by a continuous potential function $\phi:M \to \R$ which is smooth on each closed edge, and whose gradient is the current flowing in the circuit. Because $\phi$ is determined only up to adding a constant we shall normalize it by $\int_M \phi = 0.$  The energy of a state is 
$$\frac{1}{2}\int_M ||\nabla \phi ||^2 {\rm d}s,$$
and so the system can be regarded as a free massless field theory on the graph: in particular the vacuum expectation value $\langle \phi(x)\phi(y)\rangle$, when $x$ and $y$ are two nodes of the graph, is the ratio of the potential-difference $\phi(x) - \phi(y)$ to the current flowing in at $x$ and out at $y$ when no current is allowed to enter or leave at other nodes.

We encounter complex metrics when we consider a circuit in which  an alternating current with frequency $\omega$ is flowing, and in which each branch has not only a resistance $R$ but also a positive inductance $L$ and a positive capacitance $C$. In that situation the volume element $\sqrt g = R$ is replaced by the \emph{impedance} $$\sqrt g = R + {\rm i}\omega L + 1/{\rm i} \omega C,$$ 
a complex number which defines an allowable metric because Re$\sqrt g > 0$.

Quite apart from electric circuitry, however, singular one-dimensional manifolds with allowable complex metrics can arise in quantum field theory as the Gromov-Hausdorff limits of non-singular space-times of higher dimension. For example, if we embed a smooth graph $M$ in $\R^3$, then for almost all sufficiently small $\varepsilon >0$ the boundary of the $\varepsilon$-neighbourhood of $M$ is a smooth surface $M_{\varepsilon}$ whose limit is $M$ as $\varepsilon \to 0$: this is one way of viewing the passage from closed  string theory to quantum field theory.

\bs

\newpage

\no{\bf Appendix to Section 2: proof of 2.3}

\bs

If $V$ is a real vector space with an allowable complex metric then  the preceding discussion shows that it can be identified with the subspace
$$V \ = \ \exp({\rm i} \Theta /2)(\mathbb E)$$
of $\m_{\C}$. Here $\mathbb E = \R^d$ is the standard Euclidean subspace of $\m_{\C}$, and  $\Theta$ is a real diagonal matrix whose entries $\theta_1, \ldots , \theta_d$ belong to the `generalized octahedron' $\Pi_0 \subset \R^d$ consisting of those $\Theta$ whose diagonal entries $\theta_1, \ldots , \theta_d$ satisfy  the inequality (4).  We want to prove  that $\exp({\rm i} \Theta /2)$ maps each $k$-tuple ${\bf x} = \{x_1,\ldots , x_k\}$ of distinct points of $\mathbb E$ to a point of the holomorphic envelope $\hat \U_k$ of the Wightman permuted extended tube $\U_k$. In fact we shall prove the stronger statement that $\exp(\i\Theta/2)({\bf x}) \in \hat \U_k$ when $\Theta$ is a \emph{complex} diagonal matrix with Re$(\Theta) \in \Pi_0$.

The crucial fact is that $\Pi_0$ is the convex hull of its intersection $\Pi_{00}$ with the coordinate axes in $\R^d$,  (i.e.\ $\Pi_{00}$ consists of the diagonal matrices with only one entry  $\theta_r$  non-zero, and  $-\pi < \theta_r < \pi$). Our strategy is to show that  $\exp(\i \Theta /2)({\bf x}) \in \U_k$ when Re$(\Theta) \in \Pi_{00}$, and to deduce that the same is true when Re$(\Theta)$ belongs to the convex hull $\Pi_0$.  The essential tool  is Bochner's `tube theorem' ([H\"or] Thm 2.5.10), which asserts that if $P$ is a connected open subset of $\R^d$ then a holomorphic function defined in the tube domain $P\times \i\R^d$ extends holomorphically to the tube domain $P' \times \i\R^d$, where $P'$ is the convex hull of $P$.

Having fixed a $k$-tuple ${\bf x}$ in $\m_{\C}$, let us first show that if Re$(\Theta) \in \Pi_{00}$ then $\exp({\rm i} \Theta /2)({\bf x})$ is contained in $\U_k$. Suppose that the non-zero diagonal element of $\Theta$ is in the $r^{\rm th}$ place. Because $\U_k$ in invariant under the orthogonal group O$(\mathbb E)$ we can assume that the $r^{\rm th}$ basis vector $e_r$ of $\mathbb E$ is the Wick-rotated time-axis of $\m$, so that $e_r$ belongs to $\i C$, where $C$ is the forward light-cone in $\m$. With respect to the real structure $\m_{\C} = \m \oplus \i \m$ the imaginary part of the $k$-tuple $${\bf y} \ =  \ \exp(\i\Theta/2)({\bf x})$$  lies on the line $\R e_r$, and so, after ordering the points appropriately, ${\bf y}$ will belong to the forward tube in $\m_{\C}$ providing the points of ${\bf x}$ have distinct $r^{\rm th}$ coordinates. But if the $r^{\rm th}$ coordinates of Im$({\bf y})$ are not distinct, we can make them so by choosing a unit vector $e \in \mathbb E$ perpendicular to $e_r$ such that the coordinates $\langle {\bf x}, e \rangle$ are distinct, and rotating the $k$-tuple ${\bf y}$ by a small amount in the $\{e,e_r\}$-plane, again using the O$(\mathbb E)$-invariance of $\U_k$.

We now know that $\U_k$ contains an open neighbourhood of $\Pi_{00} \times \i\R^d$ in $\C^d$. To apply Bochner's  theorem we need to know that the envelope $\hat \U_k$ contains a \emph{tube}  $P \times \i\R^d$,  where $P$ is an open neighbourhood of $\Pi_{00}$ in $\R^d$.  In fact it is enough, by induction, to treat the case  $d = 2$, for that case, together with Bochner's theorem, implies that a function holomorphic in a neighbourhood of $(\Pi_0(\R^r) \cup \Pi_{00}(\R^{d-r}) \times \i\R^d$ is holomorphic in a neighbourhood of $(\Pi_0(\R^{r+1}) \cup \Pi_{00}(\R^{d-r-1})) \times \i\R^d$.

To reduce the  $d=2$ case to the standard Bochner theorem it is enough to prove the following

\bs

\no{\bf Lemma 2.8} \ {\it Let $L$ be the L-shaped subset $(\{0\} \times [0,1)) \cup ([0,1) \times \{0\})$ of the quadrant $(\R_+)^2$. Then any holomorphic function $F$ defined in a neighbourhood of $L \times \i\R^2 \subset \C^2$ can be extended  holomorphically to  $P \times \i\R^2$, where $P$ is the intersection of $(\R_+)^2$ with a neighbourhood of $L$ in $\R^2$.}

\bs

\no{\it Proof} \ For any $t \in (0,1/2)$ we define $D_t$ as the intersection of the two  unit discs $\{z \in \C: |z - (1-t)| \leq 1\}$ and $\{z \in \C: |z + (1-t)| \leq 1 \}$. Then we define $f:D_t \to \C^2$ by
$$f(z) \ = \ (-\log ((1-t)-z), -\log ((1-t) +z).$$
The map $f$ is a holomorphic embedding in a neighbourhood of $D_t$ in $\C$, and Re $f(\partial D_t)$ is contained in the coordinate axes of $\R^2$. If we choose 
 $T = (1 - {\rm e}^{-1})/2$ then Re $f(\partial D_T)$ is precisely the closure of $L$.
  
For any $\eta \in \R^2$, define $f_{\eta}:D_T \to \C^2$ by $ f_{\eta}(z) = f(z)+ \i\eta$. Then the holomorphic map $F$ is defined in a neighbourhood of the  curve $f_{\eta}(\partial D_T)$, and if we can show that $F \circ f_{\eta}$ extends holomorphically over $D_T$ then we shall have continued $F$ analytically to the tube domain $f(\mathring D_T) + \i\R^2$, and the proof will be complete.

When  a function $F$ is holomorphic in an open domain containing the boundary of a holomorphically-embedded disc  --- in this case $f_{\eta}(D_T)$ --- then to show that $F$ can be extended over the whole disc  the standard method is to show that the disc can be moved holomorphically, keeping its boundary within the domain of $F$, until the whole disc is contained in the domain of $F$; the Cauchy integral formula then defines the desired extension. In our case we can deform  $f_{\eta}(D_T)$ through the family $f_{\eta}(D_t)$ as $t$ decreases from $T$ towards 0. As $t \downarrow 0$ the domain $D_t$ shrinks to the origin in $\C$, and $f_{\eta}(D_t) \to \i\eta$, which is contained in the domain of $F$. \ \ $\spadesuit$

\newpage

\section{Quantum field theories as functors}

The traditional Wightman approach to quantum field theory is not  well-adapted to important examples such as gauge theories, especially when the space-time is not flat. Another formulation --- potentially more general --- views a $d$-dimensional field theory as something more like a group representation, except that the group is replaced by a {\it category} $\c_d^{\C}$ of space-time manifolds. The guiding principle of this approach is to preserve as much as possible of the path-integral intuition. We shall present it very briefly here, with minimal motivation.

Roughly, the objects of the category $\c_d^{\C}$ are compact smooth $(d-1)$-dimensional manifolds $\Sigma$ equipped with complex metrics $g \in {\rm Met}_{\C}(\Sigma)$. A morphism from $\Sigma_0$ to $\Sigma_1$ is a cobordism $M$ from $\Sigma_0$ to $\Sigma_1$, also with a complex metric. We shall write $M:\Sigma_0 \leadsto \Sigma_1$ to indicate a cobordism. Composition of morphisms is by concatenation of the cobordisms.  The reason for the word `roughly' is that, because there is no canonical way to give a smooth structure to the concatenation of two smooth cobordisms, we must modify the definition slightly so that an object of $\c_d^{\C}$ is not a $(d-1)$-manifold but rather is a {\it germ} of a $d$-manifold along a given $(d-1)$-manifold $\Sigma$ --- i.e. $\Sigma$ is given as a closed submanifold of a $d$-manifold $U$, but any two open neighbourhoods of $\Sigma$ in $U$ define the same object of $\c_d^{\C}$. We require $\Sigma$ to be \emph{two-sided} in $U$, and equipped with a \emph{co-orientation} which tells us which side is incoming and which is outgoing. (Nevertheless, we shall usually suppress  the thickening $U$, the co-orientation, and the complex metric $g$ from the notation.) Furthermore, two morphisms $M$ and $M'$ from $\Sigma_0$ to $\Sigma_1$ are identified if there is an isometry $M \to M'$ which is the identity on the germs $\Sigma_0$ and $\Sigma_1$. (We shall return below to the question of the existence of identity morphisms in the cobordism category.)

\bs

In terms of the category $\c_d^{\C}$ we make the 

\bs

\no{\bf Definition} \ {\it A $d$-dimensional field theory is a \emph{holomorphic} functor from $\c_d^{\C}$ to the category of Fr\'echet topological vector spaces and \emph{nuclear} (i.e. trace-class) linear maps which takes disjoint unions to tensor products.}

\bs

Unfortunately, almost every word in this definition requires further explication.

 We shall write $E_{\Sigma}$ for the vector space associated to an object $\Sigma$, and $Z_M: E_{\Sigma_0} \to E_{\Sigma_1}$ for the linear map associated to a cobordism $M: \Sigma_0 \leadsto \Sigma_1$. To say that the functor is `holomorphic' means that, for a given smooth manifold-germ $\Sigma \subset U$, the topological vector spaces $E_{\Sigma}$ form a locally trivial holomorphic vector bundle on the complex manifold Met$_{\C}(U)$ of complex metrics on $U$, and that the maps $Z_M:E_{\Sigma_0} \to E_{\Sigma_1}$ define a morphism of holomorphic vector bundles on the manifold Met$_{\C}(M)$ (to which the bundles $\{E_{\Sigma_0}\}$ and $\{E_{\Sigma_1}\}$ are pulled back).

In practice, theories are usually defined on cobordism categories where the manifolds are required to have additional structure such as an orientation or a spin-structure. These can easily be included, but are not relevant to  our account. For the same reason we do not mention that, for a theory including fermions, the vector spaces $E_{\Sigma}$ will have a mod 2 grading, and the usual sign-conventions must be applied when we speak of their tensor products.

\bs

Because our objects $\Sigma \subset U$ are really germs of $d$-manifolds, we automatically have a family of cobordisms $\Sigma' \leadsto \Sigma$ embedded in $U$, each diffeomorphic to the trivial cobordism $\Sigma \times [0,1]$ with the outgoing boundary $\Sigma \times \{1\}$ corresponding to $\Sigma \subset U$. These cobordisms can be ordered by inclusion, giving us a direct system of objects $\Sigma'$ with cobordisms to $\Sigma$. Similarly, looking downstream rather than upstream, we have a family of cobordisms $\Sigma \leadsto \Sigma''$ contained in $U$, giving us an inverse system of objects $\Sigma''$ to which $\Sigma$ maps.  For any field theory, therefore, there are natural maps
$$\lim_{\rightarrow} E_{\Sigma'} \ \  \to \ \ E_{\Sigma} \ \ \to \ \ \lim_{\leftarrow} E_{\Sigma''},$$
which we shall write
$$\check E_{\Sigma} \ \to \ E_{\Sigma} \ \to \ \hat E_{\Sigma},$$
introducing the notations $\check E_{\Sigma} = \varinjlim E_{\Sigma'}$ and $\hat E_{\Sigma} = \varprojlim E_{\Sigma''}$ for the upstream and downstream limits. 

We shall assume the functor has the \emph{continuity} property that each of these maps is injective with dense image.  The space $\hat E_{\Sigma}$, being the inverse-limit of a countable sequence of nuclear maps of Fr\'echet spaces, is a \emph{nuclear} Fr\'echet space\footnote {A very useful concise account of nuclear spaces can be found in [C].}. The other space $\check E_{\Sigma}$ is also nuclear, but  usually not metrizable: it is the dual of the nuclear Fr\'echet space $\hat E_{\Sigma^*}$, where $\Sigma^*$ denotes the germ $\Sigma$ with its co-orientation reversed. As this is such a basic point, we have included a proof as an Appendix at the end of this section. 

When we have a cobordism $M:\Sigma_0 \leadsto \Sigma_1$ we automatically get maps $\check E_{\Sigma_0} \to \check E_{\Sigma_1}$ and $\hat E_{\Sigma_0} \to \hat E_{\Sigma_1}$.
%, and both of them factorize though a map $\hat E_{\Sigma_0} \to \check E_{\Sigma_1}$. In fact this is equivalent to the original assumption that $U_M$ is nuclear, because any map from the dual of a nuclear Fr\'echet space to a nuclear Fr\'echet space is nuclear.
The space $E_{\Sigma}$ with which we began plays only a provisional role in the theory, serving to construct the fundamental nuclear spaces between which it is sandwiched.

\bs

The essential requirement we place on the functor is that it takes disjoint unions to tensor products, i.e., we are given an isomorphism of functors
$$\check E_{\Sigma} \otimes \check E_{\Sigma'} \ \to \ \check E_{\Sigma \sqcup \Sigma'} ,$$
which is associative and commutative in terms of the usual isomorphisms for the disjoint union and tensor product. There is a unique natural concept of tensor product here, because all the vector spaces are nuclear, and $\check E_{\Sigma} \otimes \check E_{\Sigma'} \cong \check E_{\Sigma \sqcup \Sigma'}$ is equivalent to $\hat E_{\Sigma} \otimes \hat E_{\Sigma'} \cong \hat E_{\Sigma \sqcup \Sigma'}$.  The functoriality means that we are assuming 
$$Z_M \otimes Z_{M'} \ = \ Z_{M \sqcup M'}$$
for two cobordisms $M$ and $M'$. 

\bs

The tensoring assumption implies that $E_{\emptyset} = \C$, where $\emptyset$ denotes the empty $(d-1)$-manifold.  Thus for a \emph{closed} $d$-manifold $M$ we have a \emph{partition function} $Z_M \in {\rm End}(E_{\emptyset}) = \C$. The whole structure of the theory is a way of expressing the sense in which the number $Z_M$ depends \emph{locally} on $M$.

\bs 

In this discussion we have still committed an abuse of language: the ``category" $\c_d^{\C}$ is not really a category because it does not have identity maps. We could deal with this by agreeing that an isomorphism $\Sigma_0 \to \Sigma_1$ is a cobordism of zero length, but then these degenerate cobordisms are represented by operators which are not nuclear. The true replacement for the missing identity operators is our assumption that the maps $\check E_{\Sigma} \to \hat E_{\Sigma}$ are injective with dense image. To avoid the abuse of language we can say that  a field theory is a functor $\Sigma \mapsto E_{\Sigma}$ from $(d-1)$-manifolds and isomorphisms to vector spaces, together with a transformation $Z_M: E_{\Sigma_0} \to E_{\Sigma_1}$ for each cobordism. Whichever line we take, we must  assume that an isomorphism $f:\Sigma_0 \to \Sigma_1$ of germs of $d$-manifolds  induces an isomorphism $f_*: E_{\Sigma_0} \to E_{\Sigma_1}$ which depends smoothly on $f$, in the sense that for any family $P \times \Sigma_0 \to \Sigma_1$ parametrized by a finite-dimensional manifold $P$ the induced map $P \times E_{\Sigma_0} \to E_{\Sigma_1}$ is smooth.

 \bs

%We shall impose one additional technical condition on the functor, namely, that the vector space $E_{\Sigma}$ depends  only on the restriction to $\Sigma$ of the metric of the germ $U$ together with a {\it finite} number of its normal derivatives along $\Sigma$. This axiom encodes the precise sense in which the notional ``path-integral" measure is local in space-time. We shall sometimes write Met$_{\C}(\hat \Sigma)$ for the space of jets of complex metrics along $\Sigma$ of the appropriate order.

\bs

Let us explain briefly 
how to get from this functorial picture to the traditional language of local observables and vacuum expectation values. For a point $x$ of a $d$-manifold $M$ we define the vector space $\O_x$ of observables at $x$ as follows. We consider the family of all closed discs $D$ smoothly embedded in $M$ which contain $x$ in the interior $\mathring D$. If $D' \subset \mathring D$ then $D \setminus \mathring D'$ is a cobordism  $\partial D' \leadsto \partial D$ and gives us a trace-class map $E_{\partial D'} \to E_{\partial D}$. We therefore have an inverse system $\{E_{\partial D}\}$ indexed by the discs $D$, and we define $\O_x$ as its inverse-limit.

Now suppose that $M$ is closed, and that $x_1, \ \ldots \  x_k$ are distinct points of $M$. Let $D_1, \ \ldots \ D_k$ be disjoint discs in $M$ with $x_i \in \mathring D_i$. Then $M' = M \setminus \bigcup \mathring D_i$ is a cobordism from $\bigsqcup \partial D_i$ to the empty $(d-1)$-manifold $\emptyset$, and defines $Z_{M'}:E_{\sqcup \partial D_i} \to E_{\emptyset} =\C$. Using the tensoring property  we can write this
$$Z_{M'} : \bigotimes E_{\partial D_i} \  \longrightarrow \ \C,$$
and then we can pass to the inverse-limits to get the expectation-value map
$$\bigotimes \O_{x_i} \ \longrightarrow \ \C.$$

We might prefer the language of ``field operators" to that of vacuum expectation values. If the space-time $M$ is a cobordism $\Sigma_0 \leadsto \Sigma_1$, then for any $x$ in the interior of $M$ --- say $x \in \mathring D \subset M$ --- the cobordisms $M \setminus \mathring D$ from $\partial D \sqcup \Sigma_0$ to $\Sigma_1$ define maps 
$$\mathcal O_x \  \to \ {\rm Hom}_{nucl}(E_{\Sigma_0}; E_{\Sigma_1}),$$
while if $x$ lies on a hypersurface $\Sigma$ an observable at $x$ defines a map $\check E_{\Sigma} \to \hat E_{\Sigma}$, i.e. it acts on $E_{\Sigma}$ as an unbounded operator. But on a Lorentzian space-time $M$ we sometimes want to make the observables at all points $x \in M$ act on a \emph{single} vector space, and to ask whether they commute when space-like separated. We shall postpone that discussion to Section 5.

\bs

One observable which we should mention is the \emph{energy-momentum tensor}. If we think of a field theory as analogous to a group representation then the energy-momentum tensor is the analogue of the induced representation of Lie algebras: for every cobordism $M:\Sigma_0 \leadsto \Sigma_1$ it is the derivative of the operator $Z_M$ with respect to the metric of $M$. This makes it a distributional symmetric tensor-density $T^{ij}$ on $\mathring M$ with values in Hom$_{nucl}(E_{\Sigma_0};E_{\Sigma_1})$. If we cover $M$ with small balls $D_i$, then by using a partition of unity we can write an infinitesimal change in the metric as the sum of contributions supported in the interiors of the $D_i$, and so the change in $Z_M$ is the sum of contributions coming from the spaces $E_{\partial D_i}$, and hence from a field operators placed at the centres of the balls $D_i$.  But to develop this picture properly needs much more discussion, which we shall not embark on here;  it probably requires the assumption that the theory is asymptotically conformal at short distances.  The case of a 2-dimensional conformal theory is treated fully in Section 9 of [Se2].

\bs

\no {\bf Lorentzian manifolds}

\bs

There is a category  $\c^{\rm Lor}_d$ which at first sight looks more relevant to quantum field theory than $\c_d^{\C}$.  Its objects are compact Riemannian manifolds of dimension $(d-1)$ and its morphisms are $d$-dimensional cobordisms equipped with real Lorentzian metrics. Fredenhagen and his coworkers (cf. [BF]) have developed the theory of quantum fields in curved space-time using a version of this category.  The category $\c^{\rm Lor}_d$ lies ``on the boundary" of the category $\c_d^{\C}$. In section 5 we shall discuss the sense in which a representation of $\c_d^{\C}$ has a ``boundary value" on $\c_d^{\rm Lor}$, at least if it is \emph{unitary}. 

\bs

\no {\bf Unitarity}

\nopagebreak

\bs

So far we have not asked for an inner product on the topological vector space $E_{\Sigma}$ associated to a $(d-1)$-manifold $\Sigma$. Our main concern in this work is with {\it unitary} theories, even though not  all interesting quantum field theories are unitary.

To define unitarity in our context, recall that,  if $\Sigma^*$ denotes the manifold germ $\Sigma$ with its co-orientation reversed, then $\check E_{\Sigma^*}$ is the dual topological vector space to $\hat E_{\Sigma}$. 
%(We obtain an isomorphism by thinking of a sequence of thinner and thinner $d$-manifolds $U$ representing the germ $\Sigma$  first as  cobordisms with both ends incoming  and then as  cobordisms with both ends outgoing.)   
Furthermore, a cobordism $M: \Sigma_0 \leadsto \Sigma_1$  can also be regarded as a cobordism from $\Sigma^*_1$ to $\Sigma^*_0$, and the two maps $E_{\Sigma_0} \to E_{\Sigma_1}$ and $E_{\Sigma_1^*} \to E_{\Sigma_0^*}$ are automatically algebraic transposes of each other. Thus $\Sigma \mapsto \Sigma^*$ is a {\it contravariant} functor.

% A pedantic-seeming terminological issue arises here. Because the $(d-1)$ manifolds $\Sigma$ which are the objects of the category $\c^{\C}_d$ are actually germs $U$ of $d$-manifolds there are {\it two} orientations present: one for $\Sigma$ and one for $U$. When the orientation of $\Sigma$ is given we orient $U$ by prescribing which side of $\Sigma$ is `incoming' and which is `outgoing'. A boundary component $\Sigma$ of a $d$-manifold $M$ can be regarded as either incoming or outgoing, and thus can acquire opposite orientations from $M$, but in both cases the $d$-dimensional orientation of $U$ is that of $M$. In our notation, both $\Sigma$  and $\Sigma^*$ have the {\it same} $d$-dimensional orientation.  If the germ $\Sigma$  describes a manifold which is getting smaller as time passes then $\Sigma^*$ will be getting bigger. 

In a unitary theory we shall not expect the vector space $E_{\Sigma}$ to have an inner product for every $(d-1)$-manifold $\Sigma$. A complex metric $g \in {\rm Met}_{\C}(\Sigma)$ has a complex conjugate $\bar g$.
If we write $\bar \Sigma$ for $\Sigma$ with the metric $\bar g$ but with its co-orientation unchanged\footnote{If our theory is defined on a category of \emph{oriented} space-time manifolds, we must give $\bar \Sigma$  the opposite orientation to $\Sigma$, although the same co-orientation in $U$.} then $\Sigma \mapsto \bar \Sigma$ is a \emph{covariant} functor. It is natural to require that there is an antilinear involution
$$E_{ \bar\Sigma} \cong \bar E_{\Sigma}. \ \ \ \ \ \ \ \ \ \ \ \ \ \ \ \ \ \ \ \ \ \ \ \ \ \ \ \ \ \ \ \ \ (5)$$
%where $\bar \Sigma$ denotes the germ with the orientation of the hypersurface $\Sigma$ reversed, but with the choice of incoming and outgoing sides unchanged. This means that $\Sigma \mapsto \bar \Sigma$ is a {\it covariant} functor, and that the contravariant functor $\Sigma \mapsto \bar \Sigma^*$ simply reverses the ``time-orientation" of the germ without doing anything to $\Sigma$ itself.

 %Ultimately, our choice of the condition (3.1) is motivated by a number of basic examples. The most basic is probably the 1-dimensional field theory describing a massive electrically-charged particle moving in a Riemannian manifold $X$ subject to no forces except a background electromagnetic field described by a hermitian line bundle $L$ on $X$ with a unitary connection $A$. The action of a trajectory defined by an oriented path $x:I=[0,1] \to X$ is given by
%$$\frac{1}{2}\int_I \ g^{-1/2}\Vert\dot x(t)\Vert^2 dt \ + \ {\rm i}  h_x(A),$$
%where $g$ is the metric of $I$ and $h_x(A)$ is the integral of the connection-form along the path $x$ (so that $\exp {\rm i} h_x(A):L_{x(0)} \to L_{x(1)}$ is the holonomy of the connection, which does not depend on the metric on the path). If the metric of $I$ is Riemannian then the first term in the action is real and positive, and does not change if we reverse the orientation of $I$; but the holonomy changes to its complex conjugate --- i.e. its inverse --- if we reverse the orientation. On the other hand if the metric of $I$ is Lorentzian then both terms in the action are pure imaginary, and both change sign when the orientation is reversed.

\bs

For a theory satisfying condition (5) the conjugate dual of the vector space $\check E_{\Sigma}$ is $\hat E_{\bar\Sigma^*}$.  We expect $\check E_{\Sigma}$ to have an inner product only when $\Sigma \cong \bar\Sigma^*$, i.e. when the $d$-manifold germ $\Sigma \subset U$ admits a reflection with fixed-point set $\Sigma$ which reverses the co-orientation and changes the metric to its complex conjugate.
%\footnote{In fact we need the diffeomorphism to preserve only the number of transversal derivatives of the metric that the vector space $E_{\Sigma}$ depends on. For free theories this number is $[(d-1)/2]$, which places no restriction at all on $\Sigma$ if $d=2$, but forces it to be a geodesic hypersurface if $d>2$.}  
Such a hypersurface-germ $\Sigma$ will be called {\it time-symmetric}. Its metric is real and Riemannian when restricted to the $(d-1)$-dimensional hypersurface $\Sigma$ itself.

\bs

We can now define a {\it unitary} theory as one which satisfies two conditions:

\bs

(i) \ \ the reality condition (5), and

\bs

(ii) \ \ {\it reflection-positivity}, in the sense that when we have a time-symmetric hypersurface $\Sigma \cong \bar \Sigma^*$ the hermitian duality between $\check E_{\Sigma}$ and $\check E_{\bar \Sigma}$ is positive-definite. 

\bs

For a unitary theory, when we have a time-symmetric germ $\Sigma$ we can complete the pre-Hilbert space $\check E_{\Sigma}$ to obtain a Hilbert space $E^{Hilb}_{\Sigma}$ with
$$\check E_{\Sigma}  \ \to \ E_{\Sigma}^{Hilb} \ \to \ \hat E_{\Sigma}.$$

%(iii) \ \ {\it boundedness}, in the sense that for every closed disc $D$ %mbedded in a closed manifold $M$ the holomorphic function $U_M: {\rm Met}_{\C}%M) \to \C$ is bounded on the sets $\mathcal U_{g,D}$ of all metrics which agree %ith a given metric $g$ except in the interior of $D$.

%\bs

%The last of these conditions is the generalization to quantum field theory of %the boundedness of the quantum mechanics propagation operator $\exp \imath t H$ %for $t$ in the upper half-plane.

\no{\bf The theory on flat tori}

\bs

The partition function of a theory on oriented flat Riemannian tori already gives us a lot of information about the theory.  The moduli space of such tori is the double-coset space$${\rm O}_d \backslash {\rm GL}_d(\R)/{\rm SL}_d(\Z) \ \cong \ Q(\R^d)/{\rm SL}_d(\Z),$$
where $Q(\R^d) = {\rm O}_d\backslash {\rm GL}_d(\R)$ is the space of positive-definite real $d \times d$ matrices. This space is an orbifold, so the partition function is best described as a smooth function $Z: Q(\R^d) \to \C$ which is invariant under SL$_d(\Z)$. Our axioms imply that $Z$ extends to a \emph{holomorphic} function
$$Q_{\C}(\R^d) \ \to \ \C,$$
but they also imply very strong constraints beyond that. Notably, each choice of a surjection $\Z^d = \pi_1(M) \to \Z$ gives us a way of writing the torus $M$ as a cobordism $\tilde M : \Sigma \leadsto \Sigma$ from a $(d-1)$-dimensional torus $\Sigma$ to itself, and then we have $Z(M) = {\rm trace} (Z_{\tilde M})$, where \mbox{$Z_{\tilde M}:E_{\Sigma} \to E_{\Sigma}$} is a nuclear operator in the vector space $E_{\Sigma}$, which is graded by the characters $\chi$ of the translation-group $T_{\Sigma}$ of $\Sigma$. More explicitly, $M$ is constructed from the product manifold $\tilde M \times [0,t]$ by attaching the ends to each other after translating by a vector $\xi \in T_{\Sigma}$, and  we  have
$$Z(A,t,\xi) \ = \ \sum_{i,\chi} n_{i,\chi }\chi(\xi) \  {\rm e}^{-\lambda_i t},$$
where $\{\lambda_i = \lambda_i(A) \}$ is the sequence (tending to $+\infty$) of eigenvalues of the Hamiltonian operator on  $E_{\Sigma}$, and the $n_{i,\chi}$ are positive integers which, for each $i$, vanish for all but finitely many characters $\chi$. 

\bs

\no{\bf Appendix  to Section 3: \ The duality $(\check E_{\Sigma})^* \ \cong \ \hat E_{\Sigma^*}$}

\bs

To keep things as general as possible, we suppose that $\Sigma \mapsto E_{\Sigma}$ is a functor from the $d$-dimensional cobordism category to a category of metrizable topological vector spaces and nuclear maps. We suppose also that the category of vector spaces is equipped with a  tensor product functor\footnote{For example, we could work with the category of Hilbert spaces with the natural Hilbert space tensor product.} which is coherently associative and commutative, and that we are given natural isomorphisms $E_{\Sigma_1} \otimes E_{\Sigma_2} \ \to \ E_{\Sigma_1 \sqcup \Sigma_2}$.

\bs

Composable cobordisms
$\Sigma_1 \leadsto \Sigma_2 \leadsto \Sigma_3$ 
  give us  maps %between nuclear Fr\'{e}chet spaces
$$E_{\Sigma_1} \to E_{\Sigma_2} \to E_{\Sigma_3}. \ \qquad  \ \qquad \ (6)$$
By reinterpreting  $\Sigma_1 \leadsto \Sigma_2$ as a cobordism $\Sigma_1 \sqcup \Sigma_2^* \leadsto \emptyset$ we get a map $E_{\Sigma_1} \otimes E_{\Sigma_2^*} \to \C$, and hence $E_{\Sigma_1} \to (E_{\Sigma_2^*})^*$. Similarly, we can reinterpret $\Sigma_2 \leadsto \Sigma_3$ as $\emptyset \leadsto \Sigma_2^* \sqcup \Sigma_3$, which gives $(E_{\Sigma_2^*})^* \to E_{\Sigma_3}$. It is easy to see that the composite $E_{\Sigma_1} \to (E_{\Sigma_2^*})^* \to E_{\Sigma_3}$ coincides with $E_{\Sigma_1} \to E_{\Sigma_2} \to E_{\Sigma_3}$. 

Yet again, performing the reinterpretations in the reverse order, we get maps
$$(E_{\Sigma^*_1})^* \to E_{\Sigma_2} \to (E_{\Sigma^*_3})^*$$
whose composite  is the transpose of the map induced by the composite cobordism $\Sigma_3^* \leadsto \Sigma_1^*$.

\bs

Now suppose that we have an infinite sequence of cobordisms 
$$\ldots \leadsto \Sigma_{i+1} \leadsto \Sigma_{i} \leadsto \Sigma_{i-1} \leadsto \ldots \ , \qquad \qquad (7)$$ indexed by $i \geq 0$, which form the downstream tail of a manifold-germ $\Sigma$, i.e. the sequence which we used above to define the space $\hat E_{\Sigma} = \lim_{\leftarrow} E_{\Sigma_i}$. 
Let us perform the two manipulations that we performed on (6) alternately on the sequence (7), thereby obtaining a sequence whose even terms are $E_{\Sigma_{2i}}$ and whose odd terms are $(E_{\Sigma^*_{2i+1}})^*$.  The inverse-limit of the whole sequence is the same as that of any cofinal subsequence. Considering the cofinal subsequence of even terms shows that the inverse-limit is $\hat E_{\Sigma}$. But the inverse-limit of the cofinal sequence of odd terms is
$$\lim_{\leftarrow} \ (E_{\Sigma_{21+1}^*})^* \ = \ (\lim_{\rightarrow} E_{\Sigma^*_{21+1}})^*.$$ This  shows that $\hat E_{\Sigma} \cong (\check E_{\Sigma^*})^*$. But, because $\hat E_{\Sigma}$ is automatically a nuclear Fr\'echet space, we can dualize again and conclude that $(\hat E_{\Sigma})^* \cong \check E_{\Sigma^*}$ also.

\bs

\section{Some analogies from representation theory}

\nopagebreak

The relation between representations of the category $\c_d^{\C}$ and of the Lorentzian category $\c_d^{\rm Lor}$ which lies ``on its boundary"  follows a pattern familiar in the representation theory of many Lie groups which have a special class of unitary representations characterized as  the boundary values of holomorphic representations of a complex semigroup by contraction operators. The essential features can all be seen in the simplest example.

\bs

The group $G = {\rm PSL}_2(\R)$ is the group of M\"obius transformations of the Riemann sphere $\Sigma = \C \cup \infty$ which map the open upper half-plane $\u$ to itself. It lies on the boundary of the complex sub-semigroup of $G_{\C} = {\rm PSL}_2(\C)$ consisting of M\"obius transformations which map the closure of $\u$ into its own interior. It is natural, however, to consider a slightly larger semigroup $G^<_{\C}$ by including the degenerate M\"obius transformations which collapse $\u$ to a single point in $\u$ --- these correspond to complex $2 \times 2$ matrices of rank one. The resulting semigroup is then a contractible open subset of the 3-dimensional complex projective space formed from the $2 \times 2$ matrices.  The topological boundary of $G^<_{\C}$ consists of the M\"obius transformations which take $\u$ to a disc or point in the upper  half-plane which touches the real axis, and the Shilov boundary consists of the group $G$ of real M\"obius transformations ---  an open solid torus --- compactified by its 2-torus boundary, which is the hyperboloid  det$(A) = 0$ in $\P^3_{\R}$ consisting of the degenerate real M\"obius transformations. (Thus the complete Shilov boundary is the part of $\P^3_{\R}$ where det$(A) \geq 0$.)

\bs

The irreducible unitary representations of the group $G = {\rm PSL}_2(\R)$ are essentially\footnote{We shall ignore the  ``supplementary" series, which is of measure zero in the space of representations.} of two kinds, the {\it principal series} and the {\it discrete series}. The best-known principal series representation is the action of $G$ on the Hilbert space of $1/2$-densities on the circle $\P^1_{\R}$ which is the boundary of $\u$ --- the general member of the series is the action on densities of complex degree $s$ with Re$(s) = 1/2$. The best-known discrete series representation is the action of $G$ on the square-summable holomorphic 1-forms on $\u$,  with the natural norm 
$$\parallel \alpha \parallel ^2 = {\rm i}\int_{\u} \alpha \wedge \bar \alpha$$
 --- more generally, for each positive integer $p$ we have the action on holomorphic $p$-forms $\alpha = f(z)(dz)^{\otimes p}$, when one must divide $\alpha \wedge \bar \alpha$ by the $(p-1)^{\rm st}$ power of the $G$-invariant area form on the Poincar\'e plane $\u$ to define the norm. 
 
 The discrete series representations obviously extend to bounded holomorphic representations of the semigroup $G^<_{\C}$ by contraction operators. They are singled out by this `positive energy' property: the principal series representations cannot extend to $G^<_{\C}$, because when $|a| < 1$ the element $w \mapsto aw$ (here $w = (z-{\rm i})/(z+{\rm i})$ is the coordinate in the unit-disc model $|w|<1$ of $\u$) of the semigroup $G^<_{\C}$ would be represented by an operator whose eigenvalues are $a^n$ for \emph{all} $n \in \Z$.  But let us notice that, though  the discrete series representations are unitary on the boundary group $G = {\rm PSL}_2(\R)$, the degenerate elements of $G^<_{\C}$, which collapse $\u$ to a point $p \in \u$, are represented by bounded operators of rank 1. So these unitary representations of PSL$_2(\R)$ do not extend unitarily to the whole Shilov boundary: the degenerate elements correspond to unbounded rank 1 operators $\xi \mapsto \langle \zeta , \xi \rangle \eta$, where $\eta$ and $\zeta$ are ``non-normalizable elements" of the Hilbert space --- i.e. they belong to an appropriate completion of it.
  
\bs 

The group $G$ is a subgroup of the group Diff$^+(S^1)$ of orientation-preserving diffeomorphisms of the circle. This infinite-dimensional Lie group does not possess a complexification, though its Lie algebra, the space of smooth vector fields on the circle, can of course be complexified. The beginning of the present work was the observation, made in the 1980s quite independently by the two authors and also by Yu. Neretin ([N], [Se1]), that there is an infinite-dimensional complex semigroup $\A$ which has exactly the same relation to Diff$^+(S^1)$ as $G_{\C}^<$ has to $G = {\rm PSL}_2(\R)$.  Its elements are complex annuli with parametrized boundary circles: one can think of them as `` exponentiations" of outward-pointing complex vector fields defined on a circle in the the complex plane. The annuli form a complex semigroup when concatenated as cobordisms, and the lowest-weight or ``positive-energy" representations of Diff$^+(S^1)$ --- and of loop groups --- which arise in 2-dimensional conformal field theory are precisely those which are boundary values of holomorphic representations of the semigroup $\A$ by trace-class operators.

\bs

The discussion of PSL$_2(\R)$ generalizes to the symplectic group $G = {\rm Sp}(V) \cong {\rm Sp}_{2n}(\R)$ of a real symplectic vector space $V$ of dimension $2n$. The role of the upper half-plane $\u$ is played by the Siegel `generalized  upper half-plane' --- the domain $\u(V)$ of positive  Lagrangian subspaces of the complexification $V_{\C}$  described in Section 1. The group $G$ lies on the boundary of a semigroup $G^<_{\C}$  which is the Siegel domain $\u(\tilde V \oplus V)$, where $\tilde V$ denotes $V$ with sign of its symplectic form reversed. A generic element of this domain is the graph of a complex symplectic transformation of $V_{\C}$ which maps the closure of $\u(V)$ into its own interior, but, just as was the case with PSL$_2(\C)$, there are degenerate elements which map $\u(V)$ non-injectively into itself.
The complex semigroup $G_{\C}^<$ has been carefully studied by Roger Howe [H], who called it the {\it oscillator semigroup}.  

The Shilov boundary of $G^<_{\C}$ is the Grassmannian of real Lagrangian subspaces of $\tilde V \oplus V$: generically, these are the graphs of elements of the real  group $G = {\rm Sp}(V)$, but this group is compactified by the addition of Lagrangian subspaces which intersect the axes of $\tilde V \oplus V$ nontrivially, and thus correspond to Lagrangian correspondences from $V$ to $V$ which are not actual maps $V \to V$. Once again, whereas Sp$^<(V_{\C})$ is a genuine semigroup, the composition-law of the real group Sp$(V)$ does not extend to the compactification.

\bs

%It may be worth mentioning that Weinstein and others (cf. [WW],[W]) have considered various ways of defining a ``category" whose objects are symplectic vector spaces and whose morphisms are arbitrary Lagrangian correspondences. Our approach provides an alternative perspective on this: when we take positive complex Lagrangians as morphisms, composition is always defined.

\bs

The group $G = {\rm Sp}(V)$ has a discrete series of unitary representations generalizing those of PSL$_2(\R)$. The most important   is the \emph{metaplectic representation} --- actually a representation of a double covering $\tilde G$ of Sp$(V)$ --- which is the action on the {\it quantization} $\H_V$ of the symplectic space $V$.  The Hilbert space $\H_V$ is characterized by the property that it contains a copy of the ray $(\bigwedge^n (W))^{\otimes (1/2)}$ for each point $W$ of the domain $\u(V)$ ---  the square-root of the natural hermitian holomorphic line bundle $\{\bigwedge^n(W)\}$ on $\u(V)$ is canonical up to multiplication by $\pm 1$, and is holomorphically embedded in $\H_V$. It is acted on by $\tilde G$ rather than $G$.

The action of $\tilde G$ on $\H_V$ is the boundary-value of a holomorphic projective representation of the oscillator semigroup $G_{\C}^<$. For $G_{\C}^<$ is just the domain $\u(\tilde V \oplus V)$, each point of which defines a ray in
$$\H_{\tilde V \oplus V} \ \cong \  \H_V^* \otimes \H_V \ \cong \  {\rm End}_{HS}(\H_V),$$
where End$_{HS}$ denotes the Hilbert-Schmidt endomorphisms.  (A more careful discussion shows that $G^<_{\C}$ is represented by operators of trace class.)

\bs

When $n = 1$ the group Sp$(V)$ is SL$_2(\R)$, a double covering of the group PSL$_2(\R)$ of M\"obius transformations we considered before. To relate the  cases of PSL$_2(\R)$ and Sp$(V)$, recall that PSL$_2(\C)$ is an open subspace of the complex projective space $\P^3_{\C}$ formed from the vector space of $2 \times 2$ matrices: in fact it is the complement of the quadric $Q^2_{\C} \cong \P^1_{\C} \times \P^1_{\C}$  defined by the vanishing of the determinant, i.e. by the matrices of rank 1.   The double covering group SL$_2(\C)$ sits inside the Grassmannian of complex Lagrangian subspaces of $\C^4$, which is a quadric 3-fold $Q^3_{\C}$ in $\P^4_{\C}$: it is a non-singular hyperplane section (corresponding to the Lagrangian condition) of the Klein quadric formed by all the lines in $\P^3(\C)$. The quadric $Q^3_{\C}$ is the branched double-covering of the projective space $\P^3_{\C}$ of $2 \times 2$ matrices, branched along the quadric $Q^2_{\C}$ of rank 1 matrices. The contractible semigroup SL$^<_2(\C)$ is the open subset of the Lagrangian Grassmannian of $\C^4$ consisting of the positive Lagrangian subspaces, and it is a double covering of PSL$^<_2(\C)$.

\bs

\section{Unitarity and global hyperbolicity}

In the previous section we saw how a holomorphic representation of a complex semigroup by contraction operators on a Hilbert space can give rise --- on passing to the boundary --- to a unitary representation of a group which is a dense open subset of the Shilov boundary of the semigroup. The remaining points of the Shilov boundary are not represented by unitary operators; the representation extends to them only in  some ``weak" sense. We now come to the analogue of this phenomenon in quantum field theory, where the Lorentzian cobordism category $\c_d^{\rm Lor}$ lies on the boundary of $\c_d^{\C}$, and the role of the open dense subgroup of the Shilov boundary is played by the subcategory of \emph{globally hyperbolic} cobordisms which we shall define below. We should  mention, however, that although the category of globally hyperbolic cobordisms is very natural, the category $\c_d^{{\rm Lor}}$ may be smaller than the optimal category we could put on the boundary of $\c_d^{\C}$. For example, the Lorentzian cobordisms could possibly be allowed to contain `black holes' surrounded by horizons, rather analogous  to the `cobordisms-with-boundaries' used to describe two-dimensional theories with both open and closed strings. We shall not pursue such speculations here.

\bs

When we have a theory defined on $\c_d^{\C}$ let us first consider how to extend the assignment $\Sigma \mapsto E_{\Sigma}$ to  a \emph{Lorentzian} germ %This is because of our assumption that $E_{\Sigma}$ depends on only finitely many derivatives of the transversal metric --- i.e. that the vector spaces $E_{\Sigma}$ form a bundle on Met$_{\C}(\hat\Sigma)$.
 $\Sigma \subset U$,  with $\Sigma$ co-oriented in $U$.  We can identify $U$ with $\Sigma \times (-\varepsilon, \varepsilon)$ by exponentiating the geodesic curves emanating perpendicularly from $\Sigma$.  The metric then takes the form $h_t - {\rm d}t^2$, where $t \mapsto h_t$ is a smooth map from $(-\varepsilon, \varepsilon)$ to the manifold of Riemannian metrics on $\Sigma$. If the germ is time-symmetric then we can define $E_{\Sigma}$ by replacing the Lorentzian metric by the `Wick rotated' Riemannian metric  $h_{{\rm i}t} + {\rm d}t^2$, which makes sense because if $h_t = h_{-t}$ then $h_t$ is a function of $t^2$, so that $h_{{\rm i}t}$ is defined and real. But this does not help for a general hypersurface, and in any case seems rather arbitrary: we shall return to this point in Remark 5.3 below.
%choose a real-analytic map $t \mapsto \tilde h_t$ with the same jet as $t \mapsto h_t$ at $t=0$, to sufficient order. This can be extended to a holomorphic map into Met$_{\C}(\Sigma)$ defined in a neighbourhood of $0 \in \C$.  We then have a family of complex metrics
%$g_{\lambda} = \tilde h_{\lambda t} + \lambda^2 {\rm d}t^2,$$
%on $\Sigma \times (-\varepsilon, \varepsilon)$ parametrized by  $\lambda$ in the open semidisc 
%$\mathring D^+ =\{\lambda \in \C: |\lambda| < 1  \  {\rm and} \ {\rm Re} \  \lambda >0\}$,
%and so we have a holomorphic vector bundle on $\mathring D^+$ with fibres $E_{\Sigma_{\lambda}}$. The bundle is constant along radial rays because the metrics on a ray are diffeomorphic, and we define $E_{\Sigma}$ for the Lorentzian germ as $E_{\Sigma_{\lambda}}$ for any positive real $\lambda$. (In (5.3) below we shall see that the bundle is canonically trivial on all of $\mathring D^+$.)

\bs

It is  less easy to assign an operator $Z_M:E_{\Sigma_0} \to E_{\Sigma_1}$ to a Lorentzian cobordism $M:\Sigma_0 \leadsto \Sigma_1$. Even if $M$ is a cylinder topologically, it can be complicated in its ``causal" structure. Consider, for example, a 2-dimensional cylindrical space-time. We saw in Section 2 that, up to a conformal multiplier, a complex metric on a surface is a pair of complex structures with opposite orientations. At the Shilov boundary the complex structures degenerate to the foliations by the left- and right-moving light-lines of a Lorentzian surface. If each light-line which sets out from the incoming boundary circle of the cylinder eventually reaches the outgoing boundary circle then each family of light-lines gives us a diffeomorphism from the incoming to the outgoing boundary. In fact (cf. [Se2] p.8 and p.16) the isomorphism classes of Lorentzian cylinders of this kind are determined up to conformal equivalence by the pair of diffeomorphisms together with a positive integer which counts the number of times that the left- and right-moving lines emanating from a given point of the incoming circle cross before hitting the outgoing circle. This agrees with the well-known fact that the Hilbert space associated to a circle in 2-dimensional conformal field theory comes with a projective unitary representation of the group Diff$^+(S^1) \times {\rm Diff}^+(S^1)$.

But the light-lines from the incoming circle can behave in a more complicated way. For example, one set of light-lines may spiral closer and closer to a closed limit-cycle of the foliation, a light-line which is a circle parallel to the incoming boundary circle of the annulus. That set of lines will then never reach the outgoing circle. One might think of this phenomenon as  akin to a black hole in the space-time, though, unlike a black hole, the Lorentzian metric here has no singularity. The ``blocked'' foliation is conformally the same as the ``degenerate annulus'' obtained by collapsing the closed light-line to a point, i.e.   a pair of discs with their centre-points identified. This is usually regarded as an ``annulus of infinite length", and it acts on an irreducible positive-energy representation of Diff$^+(S^1)$ by a projection operator of rank one, like the action of a degenerate complex M\"obius transformation  in a discrete-series representation of PSL$_2(\R)$.

\bs

In works on general relativity a Lorentzian cobordism $M:\Sigma_0 \leadsto \Sigma_1$ between Riemannian manifolds is called {\it globally hyperbolic} if every maximally-extended time-like geodesic in $M$ travels from $\Sigma_0$ to $\Sigma_1$. Such an $M$ must be diffeomorphic to $\Sigma_0 \times [0,1]$.  It is only for globally hyperbolic manifolds that, for example, the  Cauchy problem for the wave-equation on $M$ is soluble.

Of course here we are only considering {\it compact} cobordisms, which are not the usual focus in relativity theory. In the compact situation we can take the definition of global hyperbolicity to be the existence of a smooth time-function $t: M \to [0,1]$ whose gradient is everywhere in the positive light-cone,  and which is therefore a fibration with Riemannian fibres. From $t$ we obtain a diffeomorphism $M \to \Sigma_0 \times [0,1]$  by following the orthogonal trajectories to the time-slices.

The existence of a time-function on a compact Lorentzian cobordism is clearly an open condition, and so the globally hyperbolic cobordisms form an open subcategory $\c_d^{\rm gh}$ of $\c_d^{\rm Lor}$ which should play the role of the real Lie group to which the holomorphic contraction representations of Section 4 can be extended (though the result (5.2) we prove below is unfortunately weaker).

\bs

For a globally hyperbolic cobordism equipped with a time-function, the metric, in terms of the  diffeomorphism $M \to \Sigma_0 \times [0,1]$, takes  the form $h_t  + c^2{\rm d}t^2$ for some function $c: \Sigma_0 \times [0,1]\to {\rm i}\R$. A small deformation $\delta c$ of $c$ into the right half-plane  changes the Lorentzian metric into an allowable complex metric, and we could hope to define $Z_M$ in the Lorentzian case as the limit of the operators associated to such deformations. That, however, encounters the problem that the deformed metric depends not only on the choice of the deformation $\delta c$, but, more importantly, on the choice of the time-function, which should be irrelevant to the operator $U_M$. Happily, there is a better point of view, which also shows why the boundary-value of a semigroup of contraction operators is a \emph{unitary} representation. There is, after all, no obvious reason why the concatenation of a Lorentzian cobordism with its reverse should be represented by the identity operator --- quite unlike what happens with Riemannian cobordisms.  (A possible analogy is the process of making a based loop-space  into a topological group by collapsing paths which retrace their steps.) 

\bs

The passage from $\c_d^{\C}$ to $\c_d^{\rm Lor}$ is already interesting when $d=1$, i.e. for quantum mechanics rather than quantum field theory --- the case when the Euclidean path-integral can be treated by traditional measure-theory. It is worthwhile to spell out the argument in this case, before passing to higher dimensions. 

We began this work with the relation of positive energy to 1-parameter contraction semigroups. Our first task now is to understand why a holomorphic representation of the category $\c_1^{\C}$ is just such a 1-parameter semigroup, where the parameter runs through the open half-plane $\C_+ = \{z \in \C: {\rm Re}(z)>0\}$. Whereas a Riemannian structure on a closed interval is completely determined by its length, the allowable complex metrics on the interval have an infinite-dimensional moduli-space.

Any complex metric on  $I = [0,1]$ can be pulled back from the holomorphic quadratic differential ${\rm d}z^2$ on $\C$ by means of a smooth embedding $f: I \to \C$ such that $f(0) = 0$ and Re $f'(t) > 0$ for all $t \in I$. In fact the space Emb$(I;\C)$ of such embeddings is isomorphic to Met$_{\C}(I)$ {\it as a complex manifold}. If $f'(t) = 1$ when $t$ is sufficiently close to the ends of the  interval $I$ then the pulled-back metric defines a morphism $I_f: P \to P$ in the category $\c_1^{\C}$, where $P$ denotes the object defined by the germ of the standard metric on the line $\R$ at the origin. 

The crucial observation is that the operator $Z_f: E_P \to E_P$ defined by $I_f$ depends only on the point $f(1) \in \C_+$. It is as if $Z_f$ were the `contour integral' of a holomorphic differential on $\C$ along the path $f$. The argument is as follows. First, $Z_f$ does not change if $f$ is replaced by $\tilde f = f \circ \phi$ where $\phi$ is any diffeomorphism $I \to I$ which is the identity near the ends of the interval. This means that $Z_f$ does not change if $f$ moves along a curve in Emb$(I;\C)$ whose tangent vector at each point is the action of an element of the Lie algebra Vect$(\mathring I)$ of compactly supported vector fields on the interior of $I$. But then --- because $Z_f$ depends holomorphically on $f$ ---  it does not change if each tangent vector is the action of an element of the \emph{complexified} Lie algebra Vect$_{\C}(\mathring I)$. Finally, if $f,\tilde f \in {\rm Emb}(I;\C)$ define two morphisms $P \to P$ and have $f(1) = \tilde f(1)$, the tangent vectors to the obvious linear path from $f$ to $\tilde f$  are given by the action of elements of Vect$_{\C}(\mathring I)$. 

We can therefore write $Z_f = u(z)$, where $z = f(1)$. Obviously we have $u(z_1)u(z_2) = u(z_1 + z_2)$ for any $z_1,z_2 \in \C_+$. Furthermore, because the object $P$ of $\c_1^{\rm gh}$ is time-symmetric, the vector space $\check E_P$ is a pre-Hilbert space, and the unitarity condition tells us that $u(\bar z)$ is the hermitian transpose of $u(z)$.

The desired unitary semigroup $\{u({\rm i} T)\}_{T \in \R}$, which will act on the triple $\check E_P \to E_P^{Hilb} \to \hat E_P$, can now be defined as follows. As explained in Section 3, any vector $\xi \in \check E_P$ can be written $\xi = u(\varepsilon )\eta$ for some $\varepsilon > 0$ and some $\eta \in E_P$. We define $u({\rm i} T)\xi = u(\varepsilon + {\rm i}T)\eta$, which is plainly independent of $\varepsilon$. Finally, $u({\rm i} T)$ is unitary because
\begin{eqnarray*}
u(-{\rm i}T)u({\rm i}T) \xi \ &=&  \ u(-{\rm i} T)u(\varepsilon + {\rm i} T)\eta\\
&=& \ u(-{\rm i} T)u(\varepsilon/2)u(\varepsilon/2 + {\rm i} T)\eta\\
&=& \ u(\varepsilon/2 - {\rm i} T)u(\varepsilon/2 + {\rm i} T)\eta\\
&=& \ u(\varepsilon)\eta \ = \ \xi.
\end{eqnarray*}

\bs

To pass from $d=1$ to higher-dimensional cobordisms we observe that the essential step in our argument was the first case of the following 

\bs

\no{\bf Principle 5.1} \ \ \ {\it If a  $d$-dimensional cobordism $M$  is a real submanifold of a complex $d$-manifold $M_{\C}$, and $M$ has an allowable complex metric induced from a  holomorphic symmetric form $g$ on the tangent bundle $TM_{\C}$,   then the linear map $Z_M$ does not change when $M$ is moved around smoothly inside $M_{\C}$ (leaving its ends fixed), providing the restriction of $g$ to $M$ remains an allowable complex metric.}

\bs

As in the $d=1$ case, this principle holds because any infinitesimal movement of $M$ inside $M_{\C}$ is given by a complex vector field on $M$, while $Z_M$ depends holomorphically on $M$ and, being invariant under the action of \mbox{Diff$(M$ rel $\partial M)$}, does not change when $M$ moves in a direction given by the action of a complexified tangent vector to this group. 

\bs

Unfortunately, to  use  the principle we need the cobordism $M$ to be embedded in a complexification $M_{\C}$, and the only natural way to ensure this  is to pass from the smooth Lorentzian category $\c_d^{\rm Lor}$ to the corresponding \emph{real-analytic} cobordism category $\c_d^{\rm Lor,\omega}$, where both the manifolds and their metrics are assumed real-analytic. Inside this category there is the subcategory $\c_d^{\rm gh,\omega}$ of globally hyperbolic cobordisms: we shall also assume that the time-function $\tau:M \to {\rm i}[0,1]$ is real-analytic, though that could be avoided, because any smooth function can be approximated real-analytically. 

\bs

There are two ways of thinking about restricting to real-analytic cobordisms. One might think that the smooth cobordism category is the natural object, and try to eliminate the analyticity hypothesis. But one could also think that that the natural allowable space-times really do come surrounded by a thin holomorphic thickening, within which the choice of a smooth totally-real representative is essentially arbitrary. In any case, we can prove the following theorem.

\bs

\no{\bf Theorem 5.2} \ \  {\it A unitary quantum field theory as defined in Section 3 on the category $\c_d^{\C}$ induces a functor from $\c_d^{\rm gh, \omega}$ to topological vector spaces. The functor takes time-symmetric objects to Hilbert spaces, and takes cobordisms between them to unitary operators.} 

\bs

To be quite precise: the theorem asserts that if $\Sigma$ is a time-symmetric $(d-1)$-manifold germ then there is a Hilbert space $E_{\Sigma}^{Hilb}$ with
$$\check E_{\Sigma} \ \ \subset \ \ E_{\Sigma}^{Hilb} \ \ \subset \hat E_{\Sigma},$$
and a real-analytic globally hyperbolic cobordism $\Sigma_0 \leadsto \Sigma_1$ between time-symmetric hypersurfaces induces a unitary isomorphism $E_{\Sigma_0}^{Hilb} \to E_{\Sigma_1}^{Hilb}$ which also maps $\check E_{\Sigma_0}$ to $\check E_{\Sigma_1}$ and $\hat E_{\Sigma_0} $ to $\hat E_{\Sigma_1}$.

\bs

%It seems likely that the analyticity restrictions in (5.2) are unnecessary  --- we know that is true for 2-dimensional conformal theories from the representation theory of Diff$(S^1)$ --- but as the essential interest of the result is conceptual it does not seem worth pursuing the question strenuously without a significant application in mind.

\bs

\no {\it Proof of 5.2} \ \   Given  a real-analytic globally hyperbolic  cobordism  \mbox{$M:\Sigma_0 \leadsto \Sigma_1$} we choose a time function $t:M \to [0,1]$ whose level surfaces foliate $M$ by Riemannian manifolds, and, following the orthogonal trajectories to the foliation, we  identify $M$ with $\Sigma_0 \times [0,1]$ as before. 
  
  Using the real-analyticity assumptions, we can find a complexification $M_{\C}$ of $M$ to which both $t$ and $g$ can be extended holomorphically, and we can assume that $\tau = {\rm i}t:M_{\C} \to U \subset \C$ is a holomorphic fibre bundle over a neighbourhood $U$ of the interval ${\rm i} [0,1]$. Furthermore, the isomorphism $\Sigma_0 \times [0,1] \to M$ extends to a holomorphic trivialization of the bundle $M_{\C} \to U$.  For any smooth curve $f:[0,1] \to U$ such that $f(0) = 0$ and ${\rm Re} \    f'(s) > 0$ for $s \in [0,1]$ this gives us a totally real submanifold $M_f$ of $M_{\C}$ sitting over the curve. We can use the morphism associated to the cobordism $M_f$ in exactly the way we used $Z_f$ in  discussing the 1-dimensional case, to obtain a unitary operator $Z_M$ associated to the Lorentzian cobordism.
  
  It is important  that $Z_M$ does not depend on the choice of the time-function $t$ defining the foliation. For two choices of $t$ are linearly homotopic, and changing from one to the other amounts to deforming the totally-real embedding $\Sigma_0 \times [0,1] \to M_{\C}$ by a real-analytic diffeomorphism of $\Sigma_0 \times [0,1]$.
  
  \bs
 
\no{\bf Remark 5.3} \ \ \   We can apply the principle 5.1  to understand better how a theory defined on $\c_d^{\C}$ assigns a vector space $E_{\Sigma}$ to a Lorentzian germ $\Sigma \subset U$. 
 
 If the Lorentzian metric on $U$ is real-analytic then the complex theory gives us a holomorphic bundle $\{\hat E_f\}$ on the space 
%of germs of smooth embeddings  
%apparently arbitrary prescription which we used above to associate  a vector space to a time-slice in a Lorentzian manifold. The trivialization along radial lines combines with the holomorphicity of the vector bundle to give it a flat connection; but because the fibres of the bundle are infinite-dimensional this does not automatically give us a trivialization over $\mathring D^+$, and we must use a more elaborate argument.
 % First, the bundle can be extended holomorphically from $\%mathring D^+$ to  
$\mathcal J$   of germs of embeddings $f:(-\varepsilon,\varepsilon) \to \C$ such that $f(0)=0$ and Re $f'(t) >0$ for all $t$. In particular, for $\lambda \in \C_+$ we have the radial paths $f_{\lambda} \in \mathcal J$ for which $f_{\lambda}(t) = \lambda t$. But recall that $\hat E_f$ is the inverse-limit of a sequence of spaces associated to the germs of $f$ at the points $f(t_k)$, for any sequence $\{t_k \downarrow 0\}$. 

Now consider two neighbouring rays $f_{\lambda},f_{\lambda'}$ with $|\lambda| = |\lambda'|$, and choose a sequence $\{t'_k \downarrow 0 \}$ which interleaves $\{t_k \}$, i.e. $t_k > t'_k >t_{k+1}$. We can choose a path $f \in \mathcal J$ which lies in the sector bounded by the rays $f_{\lambda}$  and $f_{\lambda '}$  and coincides with them alternately in the neighbourhoods of the points $\lambda t_k$ and $\lambda't'_k$. This $f$ gives us a family of cobordisms from the germ at $\lambda' t'_k$ to the germ at $\lambda t_k$, and from the germ at $\lambda t_{k+1}$ to the germ at $\lambda 't'_k$. Putting these together, we obtain  inverse canonical isomorphisms between $\hat E_{f_{\lambda}}$ and $\hat E_{f_{\lambda '}}$. The coherence of these isomorphisms when we consider three nearby rays also follows from the principle 5.1.

By this means we see that we could have chosen \emph{any} smooth path $f$ to define $\hat E_{\Sigma}$. However the family $\hat E_f$ has the property that $\hat E_{\bar f}$ is the complex-conjugate space to $\hat E_f$, so that reversing the complex time-direction conjugates the identification of $\hat E_{\Sigma}$ with the Euclidean choice $\hat E_{f_1}$. If the Lorentzian germ $\Sigma \subset U$ is time-symmetric --- but not otherwise --- the arguments we have already used will give us a hermitian inner product on $\check E_{\Sigma}$.

\bs

\bs

\no{\bf Field operators}

\nopagebreak

\bs

Finally, we come to the Wick rotation of field operators, though our account will be sketchy. The first step is to understand how the vector space $\O_x$ of observables at a point $x$ of a space-time $M$ behaves as the metric of $M$ passes from complex to Lorentzian. We shall continue to assume that $M$ and its Lorentzian metric are real-analytic.

In Section 3 we associated a space  $\O_x$ to a germ at $x$ of a complex metric on a manifold containing $x$: it is the fibre of a bundle on the space Met$_{\C}(\hat x)$ of such germs. If we embed a Lorentzian $M$ in a complexification $M_{\C}$ there will be a holomorphic exponential map from a neighbourhood of $0$ in the complexified tangent space $T_x^{\C} = T_xM \otimes \C$ to $M_{\C}$. Inside $T_x^{\C}$ we can consider the $d$-dimensional real vector subspaces $V$ on which the metric induced  from the complex bilinear form of $T_x^{\C}$ is allowable. We saw in  (2.6) that these $V$ form a contractible open subset $\mathcal U$ of the real Grassmannian Gr$_d(T_x^{\C})$.  Exponentiating $V$ will give us a germ of a $d$-manifold with a complex metric, and hence a map $\mathcal U \to {\rm Met}_{\C}(\hat x)$. Pulling back the bundle of observables by this map gives us a bundle on $\U$, which, using  the principle (5.1) as we did in (5.3), we see to be trivial. Identifying its fibres gives us our definition of $\O_x$ for Lorentzian $M$.

\bs

We need no new ideas to see that for any Lorentzian cobordism $M:\Sigma_0 \leadsto \Sigma_1$ and any $x \in \mathring M$ an element $\psi \in \O_x$ acts as an operator $E_{\Sigma_0} \to E_{\Sigma_1}$. Furthermore, if $x$ lies on a time-slice $\Sigma$ we get an operator $\psi \in {\rm Hom}( \check E_{\Sigma}; \hat E_{\Sigma})$, i.e. an unbounded operator in $E_{\Sigma}$, simply by considering the cobordisms corresponding to a sequence of successively thinner collars of $\Sigma$. Indeed the same argument shows that if $x_1, \ldots , x_k$ are distinct points on $\Sigma$,  we have a map
$$\O_{x_1} \otimes \ldots \otimes \O_{x_k} \ \ \to \ \  {\rm Hom}( \check E_{\Sigma}; \hat E_{\Sigma})$$
which does not depend on choosing an ordering of the points. 

\bs

In the introduction we  mentioned the Wightman axiom  that field operators at space-like separated points must commute. We can now see how this follows from our framework, at least in a globally hyperbolic space-time. For the spaces $\check E_{\Sigma_t} \subset E_{\Sigma_t}^{Hilb} \subset E_{\Sigma_t}$ for all  times $t_0 \leq t \leq t_1$ can be identified with those at time $t_0$ by the unitary propagation $Z_{t,t'}$ from time $t$ to a later time $t'$ to get a single rigged Hilbert space $\check E \subset E^{Hilb} \subset \hat E$, and we can define an unbounded operator
$$\tilde \psi \ \ = \ \ Z_{t_0,t}^{-1}\circ \psi \circ Z_{t_0,t}:\check E \to \hat E$$
for any $\psi \in \O_x$ with $x \in \Sigma_t$. Furthermore, if we change the choice of time-function on the cobordism, so that $x$ lies on a different time-slice, then $\tilde \psi$ will not change.

The fact that two observables $\psi,\psi'$ situated at space-like separated points $x,x'$ give rise to operators $\tilde \psi, \tilde \psi'$ which are composable, and commute, is now clear. For if $x$ and $x'$ are space-like separated we can choose a single time-slice $\Sigma_t$ which contains them both, and we see that the composed operator, in either order, is  $Z_{t_0,t}^{-1}\circ ( \psi \otimes \psi') \circ Z_{t_0,t}$.

\bs

\no{\bf The domain of holomorphicity of the vacuum expectation values}

\bs

We end with a conjecture about a question arising in the traditional treatment of field theories defined in the standard Minkowski space $\m = \R^{d-1,1}$. There, the Wightman axioms imply that the vacuum expectation values, initially  defined as distributions or other generalized functions on the $k$-fold products $\m \times \ldots \times \m$, are  boundary values of holomorphic functions defined in an open domain $\U_k$ in the complexified space $\m_{\C}\times \ldots \times \m_{\C}$. The definition of $\U_k$,  known as the `permuted extended tube',  was given in Section 2. Recall that $\U_2$ consits of all pairs of points $x,y$ such that $||x-y||^2$ is not real and $\leq 0$.
% --- constructed from the Minkowski geometry\footnote{A set of points $x_1, \ldots , x_k$ belongs to $\mathcal U_k$ if, after ordering them suitably, there is an element $\gamma$ of the complexified Lorentz group such that the imaginary part of $\gamma(x_i - x_{i+1})$ belongs to the forward light-cone for each $i$. }. If $k=2$, for example, $\mathcal U_k$ consists of all $(x_1,x_2) \in \m_{\C}\times \m_{\C}$ such that $||x_1 -x_2||^2$  is not real and negative, where $|| \ \ ||^2$ denotes the complex-bilinear extension of the Minkowski inner product (using the normalization for which $||\xi||^2 < 0$ when $\xi$ is time-like).

If $k > 2$, however,  $\mathcal U_k$ is known not to be holomorphically convex, so it cannot be the largest complex manifold to which the expectation values can be analytically continued. It is an old problem to describe this largest manifold $\V_k$, or even the  holomorphic envelope of $ \mathcal U_k$.

The ideas of this paper suggest  a candidate for $\V_k$. It sits over the open subset $\check \V_k$ of all $k$-tuples ${\bf x} = \{x_1, \ldots ,x_k\}$ of distinct points in $\m_{\C}$ which lie on some totally-real submanifold $M$ with two properties:

(i) \ the metric on $M$ induced from $\m_{\C}$ is allowable, and

(ii) \ $M$ projects surjectively onto the usual real Euclidean subspace $\mathbb E = \R^d$ of $\m_{\C} = \mathbb E \oplus \i\mathbb E$.

Notice that, by the remark before Prop.\ 2.6, the projection $M \to \mathbb E$ is a local diffeomorphism if the metric of $M$ is allowable, so (ii) implies that $M$ is the graph of a smooth map $\mathbb E \to \i\mathbb E$.

Let $\F_k$ denote the space of all pairs $(M,{\bf x})$ satisfying the above conditions. It is an infinite-dimensional complex manifold projecting to the open subset $\check \V_k$ of Conf$_k(\m_{\C})$, and it is easy to see that the map $\pi: \F_k \to \check \V_k$ is open.  We define $\V_k$  as the largest Hausdorff quotient manifold of $\F_k$ through which $\pi$ factorizes and which maps to $\check V_k$ by a local diffeomorphism. Thus two points $(M,{\bf x}),\, ( M',\bf x)$ of the fibre $\F_{k,\bf x}$  of $\pi$ at $\bf x$ have the same image in $\V_k$ if they are in the same connected component of the fibre, but --- as that equivalence relation need not give a Hausdorff quotient --- also if there are paths $\gamma,\gamma'$ from $(M,\bf x)$ and $(M',\bf x)$ to a third point
$(M'',\bf x'')$ of $\F_k$ which cover the same path from $\bf x$ to $\bf x''$ in $\check V_k$.

\bs

To motivate the  definition of $\V_k$ we must enlarge our framework to allow Lorentzian space-times whose time-slices are not compact. The simplest way to do this is to introduce the cobordism category in which a morphism is the part of a  $d$-dimensional allowable submanifold $M$ of $\m_{\C}$ cut off between two time-symmetric hypersurfaces.

A field theory defined and holomorphic on this category, if it has a Lorentz-invariant vacuum state in a natural sense, will have vacuum expectation values which are holomorphic functions $\E_k$ on the spaces $\F_k$ of  pairs $(M,{\bf x})$.  Strictly speaking, $\E_k$ is a holomorphic section of a bundle on $\F_k$, but we can use the local diffeomorphism $M \to \mathbb E$ to trivialize the bundle, giving us a holomorphic function
$$\mathcal E_k: \mathcal F_k \  \to  \ {\rm Hom}(\O^{\otimes k};\C),$$
where $\O$ is the space of observables at a point of $\mathbb E$.

Our much-used Principle 5.1 tells us that the value of the function $\mathcal E_k$ does not change if, while holding the marked points {\bf x} fixed in $\m_{\C}$, we move $M$ smoothly in the allowable class. So in fact we have a holomorphic function on  $\F_k$ which is constant on the connected components  of the fibres of $\F_{k,\bf x}$ of the map $\pi$, i.e. to the isotopy classes of allowable manifolds conatining $\bf x$.

\bs  

Unfortunately we have no proof that $\V_k$ is a domain of holomorphy, but  at least we can assert

\bs

\no {\bf Proposition 5.4} \ {\it $\V_k$ contains the Wightman domain $\U_k$.}

\bs

Furthermore, we saw in Proposition 2.3 that the holomorphic envelope of $\U_k$ contains $\V_k^{flat}$, the part of $\V_k$ represented by flat affine submanifolds $M \subset \m_{\C}$.

\bs

\no{\it Proof of 5.4} \ First, $\V_k$ is invariant under the complex orthogonal group of $\m_{\C}$, and under reorderings of the points ${\bf x} = \{x_1, \ldots ,x_k\}$. So it is enough to consider ${\bf x}$ such that the imaginary part of $x_{i+1} - x_i$ belongs to the forward light-cone $C \subset \m$ for each $i$.

Smoothing the obvious polygonal path joing the points, we can thus assume that the $x_i$ lie on a curve $x: \R \to \m_{\C}$ whose derivative  Im$(x'(t))$ belongs to $C$ for all $t$. But then we can choose, smoothly in  $t$, a set of $d-1$ orthonormal vectors $e_j(t)$ in $\m_{\C}$ which are all orthogonal to $x'(t)$. Let $M_t$ be the \emph{real} vector subspace of $\m_{\C}$ spanned by the vectors $e_j(t)$. The points ${\bf x}$ lie on the $d$-dimensional real ruled manifold $M$ swept out by the  affine $d-1$-planes  $x(t)+ M_t$, and the metric of $M$ is clearly allowable. \ \ $\spadesuit$

\bs

\no{\bf References}

\bs

\begin{enumerate}
\item[[BF]] Brunetti, R.,   and K. Fredenhagen, \,{\it Quantum field theory on curved backgrounds}. \, Lecture Notes in Physics {\bf 786}, 129 -- 155, Springer 2009.
\item[[C]] Costello, Kevin, \, {\it Renormalization and effective field theory}. \, Math. Surveys and Monographs {\bf 170}, Amer.\ Math.\ Soc. 2011.
\item[[F]] Fermi, E., \,{\it Quantum theory of radiation}. \,Rev.\ Mod.\ Phys, {\bf 4}\, (1932), 87--132.
\item[[FH]] Feynman, R. P. and A. R. Hibbs, \,{\it Quantum mechanics and path integrals.} \, McGraw-Hill, 1965.
\item[[H\"or]] H\"ormander, Lars, \, {\it An introduction to complex analysis in several variables}. \, North-Holland, 1990. 
\item[[H]] Howe, Roger, \, {\it The Oscillator Semigroup}. \, Proc.\ Symp.\ Pure Math. {\bf 48}, 1196 -- 1200, \, Amer.\ Math.\ Soc. 1988.
\item[[Ka]] Kazhdan, David, \, {\it Introduction to QFT.} \, In {\it Quantum fields and strings: a course for mathematicians.} Vol 1 (Princeton 1996/1997) 377--418. Amer.\ Math.\ Soc. 1999.
\item[[Ke]] Kelnhofer, Gerald, \, {\it Functional integration and gauge ambiguities in generalized abelian gauge theories.} \, J.\ Geom.\ Physics {\bf 59} (2009) \, 1017--1035. (arXiv:0711.4085 [hep-th])
\item[[N]] Neretin, Yu. A., \, {\it Holomorphic continuations of representations of the group of diffeomorphisms of the circle.} \, Mat. Sb. {\bf 180} \, (1989), 635--57. \ (English translation Math.\ USSR-Sb. {\bf 67} (1990).
\item[[PS]] Pressley, A., and G. Segal, \, {\it Loop Groups}. Oxford U.P. 1986.
\item[[Se1]] Segal, Graeme, \, {\it The definition of conformal field theory}. \, In: {\it Differential geometrical methods in theoretical physics. (Como 1987)}\, NATO Adv.\ Sci.\ Inst. Ser C, Math Phys. Sci. {\bf 250}, 165 -- 171, Kluwer 1988.
\item[[Se2]] Segal, Graeme, \, {\it The definition of conformal field theory.} \, In: {\it Topology, Geometry, and Conformal Field Theory}, \, ed. U. Tillmann, \, London Math.\ Soc. Lecture Notes {\bf 308} (2004), 421 -- 577.
\item[[Sz]] Szabo, R., \, {\it Quantization of higher abelian gauge theory in generalized differential cohomology.} \, In: {\it Proc. 7$^{\rm th}$ Internat.\ Conf.\ on Math.\ Methods in Physics} \, (ICMP2012)  (arXiv:1209.2530 [hep-th]).
\item[[SW]] Streater, R. F.,  and A. S. Wightman, \, {\it PCT, Spin and Statistics, and all that.} \, Princeton U.P. 2000. \ ($1^{\rm st}$ edn Benjamin 1964)
%g\item[[V]] Vladimirov, V. S., \, {\it Methods of the theory of functions of many complex variables.} The MIT Press, Cambridge, Mass., 1968. (Russian original: Nauka Press, Moscow 1964.)
\item[[WW]] Weinberg, S., and E. Witten, \, {\it  Limits on massless particles.} \, Phys. Lett. B{\bf 96} (1980), 59 -- 62.

\end{enumerate}

\

\textit{E-mail addresses}: \texttt{maxim@ihes.fr},  \texttt{ graeme.segal@all-souls.ox.ac.uk}

\end{document}